\documentclass[5p,times]{elsarticle}

\usepackage[hidelinks]{hyperref}
\usepackage{cleveref}
\usepackage{adjustbox}
\usepackage{booktabs}
\usepackage[group-separator={,}]{siunitx}
\usepackage{pifont}
\usepackage{makecell}

\crefformat{section}{\S#2#1#3}
\crefformat{subsection}{\S#2#1#3}
\crefformat{subsubsection}{\S#2#1#3}
\crefrangeformat{section}{\S\S#3#1#4 to~#5#2#6}
\crefmultiformat{section}{\S\S#2#1#3}{ and~#2#1#3}{, #2#1#3}{ and~#2#1#3}

\newcommand{\cmark}{\ding{51}}%
\newcommand{\xmark}{\ding{55}}%

\journal{Future Generation Computer Systems}

\graphicspath{{figures/}}

\begin{document}
	
\clearpage
\thispagestyle{empty}

\begin{center}
\noindent\fbox{%
	\parbox{\textwidth}{%
		\vspace{40pt}
		\Large \centering
		CC-BY 4.0
		\\
		This is the author's pre-print version of article \textbf{``Transparent Serverless execution of Python multiprocessing applications''} published in journal \textbf{Future Generation Computer Systems} (Volume 140, March 2023, Pages 436-449).
		\\		
		\texttt{DOI: \href{https://doi.org/10.1016/j.future.2022.10.038}{10.1016/j.future.2022.10.038}}\vspace{10pt}\vspace{40pt}
	}%
}
\end{center}
\clearpage

\begin{frontmatter}

\title{Transparent Serverless execution of Python multiprocessing applications}

\author[urv]{Aitor Arjona}
\ead{aitor.arjona@urv.cat}
\author[urv]{Gerard Finol}
\ead{gerard.finol@urv.cat}
\author[urv]{Pedro~Garc\'ia L\'opez}
\ead{pedro.garcia@urv.cat}

\address[urv]{Universitat Rovira i Virgili, Tarragona, Spain}

\begin{abstract}
Access transparency means that both local and remote resources are accessed using identical operations. With transparency, unmodified single-machine applications could run over disaggregated compute, storage, and memory resources. Hiding the complexity of distributed systems through transparency would have great benefits, like scaling-out local-parallel scientific applications over flexible disaggregated resources in the Cloud.

This paper presents a performance evaluation where we assess the feasibility of access transparency over state-of-the-art Cloud disaggregated resources for Python multiprocessing applications. We have interfaced the multiprocessing module with an implementation that transparently runs processes on serverless functions and uses an in-memory data store for shared state.

To evaluate transparency, we run in the Cloud four unmodified applications: Uber Research's Evolution Strategies, Baselines-AI’s Proximal Policy Optimization, Pandaral·lel’s dataframe, and ScikitLearn’s Hyperparameter tuning. We compare execution time and scalability of the same application running over disaggregated resources using our library, with the single-machine Python multiprocessing libraries in a large VM. For equal resources, applications efficiently using message-passing abstractions achieve comparable results despite the significant overheads of remote communication. Other shared-memory intensive applications do not perform due to high remote memory latency.

The results show that Python's multiprocessing library design is an enabler towards transparency: legacy applications using efficient disaggregated abstractions can transparently scale beyond VM limited resources for increased parallelism without changing the underlying code or architecture.
\end{abstract}


\end{frontmatter}


\section{Introduction}

Coulouris et al. \cite{transparency_coulouris} define transparency as ``the concealment from the user and application programmer of the complexities of distributed systems''. Access transparency allows to execute unmodified parallel code in a distributed environment, where resources (CPUs, memory) are distributed over many machines but accessed as if they were arranged in a single local machine.

The motivation of this paper is very simple: Low Latency $\Rightarrow$ Disaggregation $\Rightarrow$ Full Transparency. The downward trend in network latency \cite{rpcs_general_fast, time_for_low_latency} suggests that resource disaggregation is increasingly viable, which provides the opportunity to achieve access transparency in the next years \cite{serverless_endgame, disaggregation, disaggregation_and_app}. Resource disaggregation in the Cloud has been the key to flexible scaling models provided by current serverless services such as Function-as-a-Service (FaaS) or Object Storage. Serverless services have also proven to be effective for massive parallel computing applications \cite{pywren, numpywren} and even Big Data processing at scale \cite{excamera}.

If we can use identical operations for both local and remote resources with no significant performance degradation, it is then possible to unify the local and remote programming paradigms. In such a case, developer productivity would be greatly increased, as transparency would facilitate and make more accessible to program a distributed system, making the use of specific middleware redundant. Moreover, we could port legacy monolithic applications and scale them on a Cloud setting with flexible resources. New iterations of software modernization or architecture re-engineering could be avoided, saving maintenance costs and time to engineers. With transparency, resources could be adapted for compute-intensive applications to process larger workloads than a single machine could withstand without having to modify the underlying code or architecture.

As a matter of fact, in the European project Horizon 2020 Cloudbutton\footnote{\href{https://cloudbuton.eu}{https://cloudbutton.eu}}, we aim to overly simplify and democratize the use of the Cloud for scientific computing using serverless technologies. Cloudbutton's core objective was inspired by a professor of computer graphics at UC Berkeley that wondered ``Why is there no cloud button?'' \cite{pywren}. His students wished they could just ``push a button'' and have their existing single-machine code running on the Cloud. Therefore, transparency is an important consideration, because many of the tools currently used by data scientists are legacy applications that are difficult to parallelize or move to the Cloud. Moreover, data scientists are unlikely to be knowledgeable about Cloud technologies. Achieving access transparency could effortlessly unleash the potential of Cloud flexible and serverless resources to effectively speed up scientific computing pipelines.

Nonetheless, the distributed systems community has consistently criticized the idea of transparency. The reason is that Distributed Shared Memories (DSM), studied in depth in the past \cite{dsm_survey}, suffered from complexity and performance issues, which made achieving transparency difficult. Waldo et al. \cite{note_on_distsys} already discussed in 1994 that, in the context of Object Oriented Programming, the usage of remote objects as if they were local is incorrect and leads to performance issues and general unreliability. Although the authors did not explicitly discuss access transparency in their article, their conclusions suggest that remote memory access latency and partial failures make full transparency unfeasible.

The main criticism of transparency is that remote memory will never be as fast as local memory. Nevertheless, not all parallel programming models require intensive access to shared memory. On the contrary, many parallel applications just rely on communication and synchronization primitives that could be efficiently disaggregated.

\begin{figure}[!htb]
	\centering
	\includegraphics[width=0.35\textwidth]{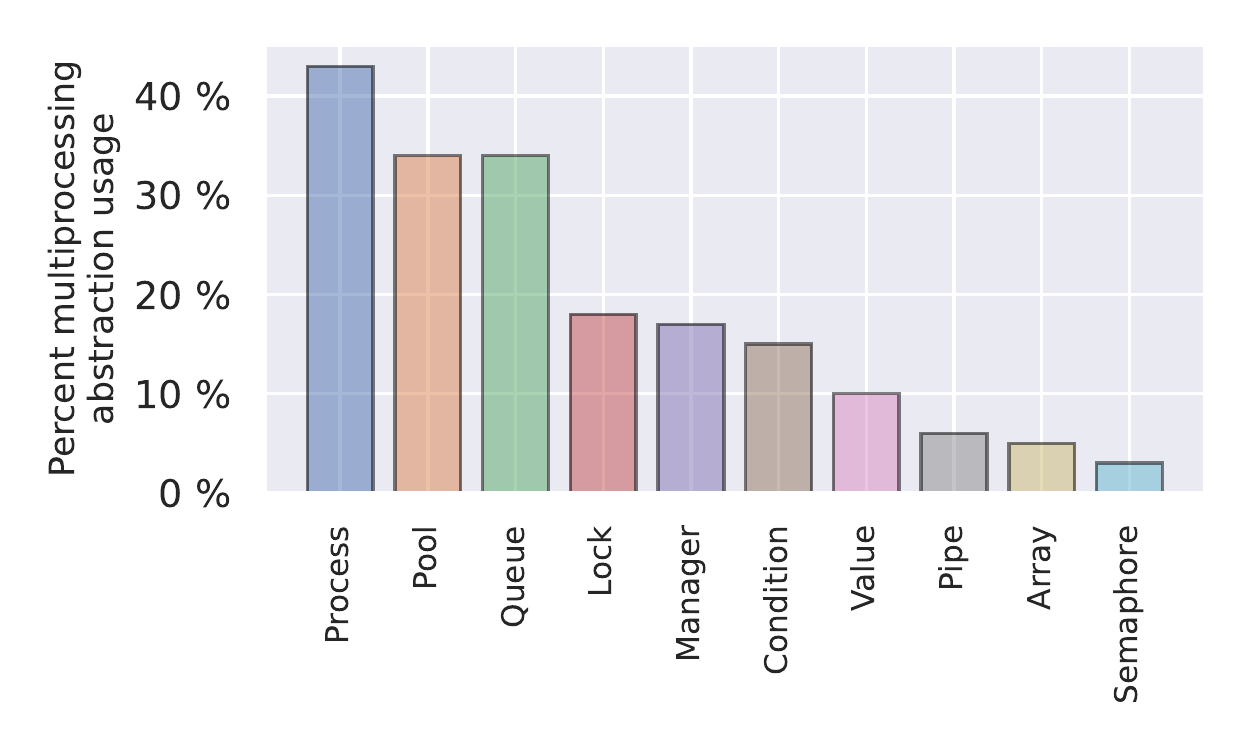}
	\caption{Percentage of usage of the main multiprocessing abstractions of the top 100 most starred GitHub Python repositories that use multiprocessing.}
	\label{fig:github}
\end{figure}

In this line, the Python programming language uses processes to achieve true parallelism. Shared state in Python multiprocessing mainly consists of message passing (Pipes, Queues) or remote calls to other processes (Managers). To reinforce this point, we have analyzed the top $100$ most starred repositories on GitHub that use Python multiprocessing and found out that Queues and Managers are the most used abstractions for state sharing (Figure \ref{fig:github}). In this aspect, the problems and limitations caused by DSMs would not apply, thus transparently adapting multiprocessing Python applications to a distributed environment is more feasible.

This article presents a performance study to evaluate if the inherent scalability of serverless functions, together with a disaggregated and consistent in-memory storage component, enables to transparently run unmodified Python multiprocessing applications over disaggregated serverless compute resources at scale. We want to run the same application with the same workload both in a VM (Virtual Machine) on AWS EC2 and with serverless functions on AWS Lambda, in order to compare execution time, speedup, parallelism and to determine the possible overheads originated by moving to a distributed environment.

For this purpose, we have extended the Lithops serverless computing framework \cite{lithops} with a module that fully implements the Python multiprocessing interface. This re-implementation leverages serverless functions for processes and Redis database for stateful multiprocessing abstractions (shared state, queues, locks\dots). Python applications written with the multiprocessing library can then be transparently ported to the Cloud by only changing the \texttt{import} statement.

For the performance study, we have used four scientific applications that make use of Python multiprocessing for parallelism: Evolution Strategies, Proximal Policy Optimization, Scikit-Learn Grid Search and Pandas dataframes. For each of them, we have compared its local execution on VMs currently available in AWS EC2 with its equivalent execution on AWS Lambda, while maintaining the same code, to asses if the application could be further scaled using serverless flexible resources despite the limitations and overheads generated.



The results show that it is actually possible to effectively run a local-parallel Python application using serverless resources, however, the application must efficiently use message-passing multiprocessing abstractions (Pipes, Queues, Managers\dots) in order to perform. On the contrary, applications that heavily use shared memory abstractions (Arrays) struggle to achieve competent results due to the high latency of remote memory. Yet, legacy applications that use efficiently disaggregated communication and synchronization abstractions yield good performance, opening the possibility to effortlessly scale on Cloud's flexible resources.

Our contributions are:
\begin{enumerate}
	\item Performance study and evaluation of access transparency feasibility by comparing the execution of 4 unmodified applications (Evolution Strategies, Proximal Policy Optimization, Scikit-Learn Grid Search and Pandas data-frames) on a single VM and over serverless functions to analyze the speedup, parallelism and overheads.
	
	\item We outline key insights from this study: (i) it is possible to run and scale unmodified Python multiprocessing applications with minor degradation compared to a VM, (ii) existing serverless overheads were partially masked by Hyper-Threading inefficiencies observed in a single VM, and (iii) Python's multiprocessing stateful abstractions clearly facilitated transparency (other languages directly accessing memory references or pointers would be more complex to intercept).
	
	\item Extension to the Lithops computing framework with a module that implements the Python multiprocessing library and allows to execute remote processes in serverless functions in a transparent way. The project is open source and the source code is publicly available on Git-Hub\footnote{\href{https://github.com/lithops-cloud/lithops}{https://github.com/lithops-cloud/lithops}}.
\end{enumerate}

The open questions presented in this article are: Can we run legacy single-machine parallel applications in the Cloud and scale them transparently using serverless resources? Can we program the Cloud as an infinite multi-core machine?


\section{Related work}



\noindent\textbf{OS-level transparency.} Hiding the complexity of distributed systems is a recurring topic in the systems field. Recent industrial trends on Disaggregated Data Centers (DDC) \cite{disaggregation} advocate for a distributed OS transparently leveraging disaggregated hardware resources like processing, memory or storage.

For example, LegoOS \cite{legoOS} is a disaggregated OS that implements a subset of the Linux system call interface so that existing unmodified Linux applications can run on top of it. LegoOS shows how two unmodified applications can be run in a distributed way: Phoenix (a single-node multi-threaded implementation of MapReduce) and TensorFlow. LegoOS is however not demonstrating scalability or complex scenarios involving mutable memory. Even worse, implementing the entire Linux APIs over disaggregated resources is a daunting engineering task.

Another approach in access transparency at the operating system layer is GiantVM \cite{giantVM}. GiantVM uses virtualization to run an unmodified guest OS over a distributed cluster. In contrast to the traditional many-to-one virtualization paradigm (running multiple OS in one machine), GiantVM implements one-to-many virtualization (running a single OS in many machines). GiantVM uses \textit{Infrastructure as a Service} (IaaS) to run the distributed OS over a cluster in a Cloud setting. Approaches based on current Cloud technologies are more feasible at the moment, since DDCs are not yet available to the general public.

In \cite{disaggregation_and_app}, the authors propose to augment operating systems for disaggregation, by exposing explicitly the disaggregated resources to applications and thus opening efficient and optimized co-designs between applications and the remote resources.

\medskip

\noindent\textbf{Application-level transparency.} Instead of at the operating system level, transparency can also be achieved easily at the application level. If the interface that is used by the application to access local resources is replaced or wrapped by another implementation that accesses disaggregated resources instead, we could then transparently run unmodified local code in a distributed fashion.

Early efforts in the Serverless community propose \textit{FaaSification} as an automated process to move existing code to Serverless Functions \cite{spillner2017transformation}. Although this only implied simple functional code, and not entire applications.

Fiber \cite{fiber} is a library that implements Python's multiprocessing API to run remote processes in a distributed Kubernetes cluster. In their article, they execute several stateful AI applications that are programmed for local parallel execution using Python's multiprocessing library. By replacing multiprocessing with Fiber, those unmodified applications can transparently scale and exploit parallelism on a distributed Kubernetes cluster.

Although our work is closely related to Fiber, they compare Fiber with other distributed computing frameworks, while we wanted to focus on studying access transparency and to find the differences between local and distributed execution exploiting the high scalability of FaaS. Moreover, Fiber does not implement all Python multiprocessing abstractions (for example, it misses the \texttt{Lock}).

Another example of transparency at the application level is Crucial \cite{crucial}. Crucial is a library that implements Java's threading interface which allows threads to be transparently executed as serverless functions. Using serverless, Crucial highly improves scalability thanks to the inherent elasticity of \textit{FaaS}. Crucial also provides stateful abstractions based on distributed shared objects that reside in a disaggregated in-memory layer (Infinispan). Nevertheless, Crucial is not offering transparency to Java applications, since it offers an explicit programming model for shared remote objects.

Kappa \cite{kappa} is a serverless computing framework that focuses on providing fault tolerance for stateful serverless applications by means of checkpointing mechanisms and continuation functions. Contrary to our work, they do not emphasize access transparency in their contributions. Yet, they claim that their framework requires minimal modifications to the original code, because Kappa's static code analysis system is able to create checkpoints automatically. However, a special API is required to invoke tasks (\texttt{spawn} and \texttt{spawn\textunderscore map}) and to pass messages between functions. We believe it would be interesting to study fault-tolerant access transparency with Kappa framework in future work.


We differ from related work in that we are the first work that evaluates full transparency over serverless disaggregated resources intercepting multi-processing parallel libraries.





\section{Enabling transparency for Python multiprocessing}

\begin{figure*}[htbp]
	\centerline{\includegraphics{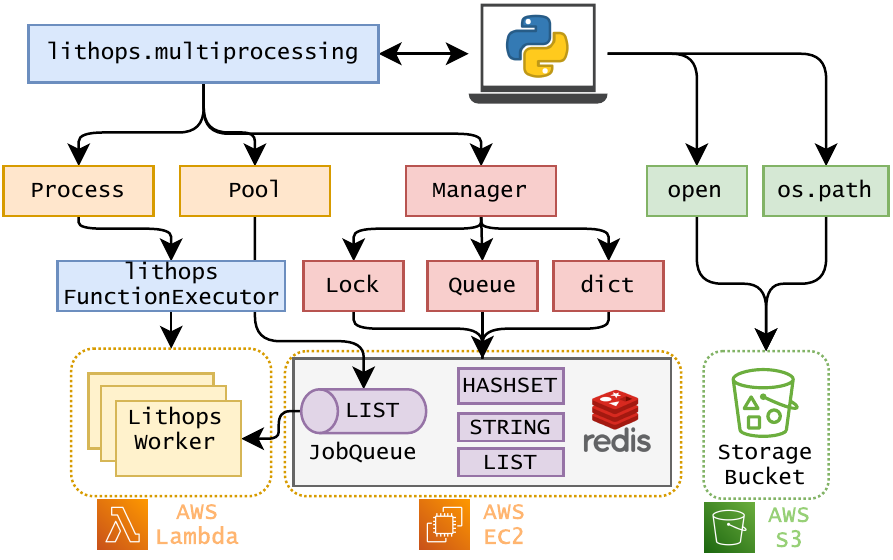}}
	\caption{Architecture diagram.}
	\label{fig:arch}
\end{figure*}

This section describes how we achieved access transparency to serverless and disaggregated resources through the re-im\-plementation of the Python multiprocessing library. Figure \ref{fig:arch} depicts a general overview of the architecture.

We have leveraged the Lithops framework \cite{lithops} to execute parallel local applications over disaggregated serverless functions. Lithops enables the execution of local serial code to be run over massively parallel serverless functions. Lithops acts as an abstraction layer that simplifies the exploitation of the main FaaS services present in public clouds for highly parallel tasks. One of Lithops design principles is to ensure portability between clouds. The same application can be seamlessly ported from one cloud provider to another, which prevents vendor lock-in.

We have extended Lithops with a \texttt{multiprocessing} module which implements in its entirety the original Python \texttt{multi\-processing} interface. Computation abstractions (like \texttt{Process} and \texttt{Pool}) use Lithops \textit{FunctionExecutor} API. Inter-process Comunication (IPC) and synchronization abstractions (like \texttt{Lock} and \texttt{Queue}) are implemented using Redis key-value in-memory database.

\subsection{Disaggregated compute resources}
\label{sec:lithops_arch}

\subsubsection{Lithops workflow}

\begin{figure}[htbp]
	\centerline{\includegraphics{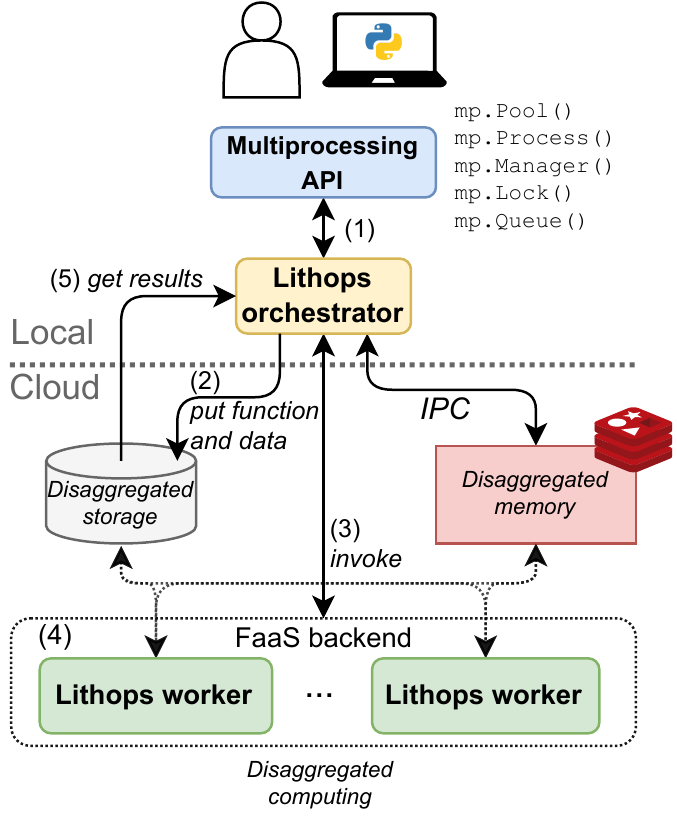}}
	\caption{Lithops operation workflow.}
	\label{fig:lithops_workflow}
\end{figure}

Lithops follows a main/worker architecture where a local process acts as the orchestrator and coordinator of the workers that are deployed and executed in the FaaS backend as serverless functions. A diagram of the general operation of Lithops is shown in Figure \ref{fig:lithops_workflow}.

First, the user interacts with Lithops multiprocessing API, which is a wrapper around Lithops framework API (1). Lithops automatically detects, serializes and uploads to storage the processes' dependencies, process function code and input arguments (2). Next, the Lithops orchestrator invokes the corresponding number of serverless functions against the FaaS backend (3). For example, every \texttt{Process} corresponds to a single function. The Lithops worker (4) is a generic serverless function that handles the execution of Lithops job tasks. It downloads the previously uploaded code, data and dependencies from storage, deserializes them and it executes the user's function in a wrapper that handles errors. When the task is completed, the result is uploaded back to storage. The Lithops orchestrator synchronizes completed tasks by pulling the contents of the storage (5) -- a task has finished when the result key is listed. Finally, the results are downloaded and returned to the parent application.

\subsubsection{Serverless job queue}

Forking many local processes for very fine-granular and short-running tasks is slow, and is much worse in Lithops, as the overhead of invoking many functions can be prohibitively expensive. For this reason, Python multiprocessing implements the \texttt{Pool} abstraction. A \texttt{Pool} represents a fixed-size group of processes that are forked at the time of the pool instantiation. It has methods like \texttt{starmap()}, \texttt{apply\textunderscore async()} or \texttt{map\textunderscore async()} which allows to offload tasks to the worker processes at a higher level, instead of manually creating and managing \texttt{Process} instances. In a process pool, each operation (\textit{map}, \textit{apply\textunderscore async}\dots) creates one or more jobs, which are enqueued to a \textit{job queue}. The worker processes get and execute jobs from the queue. This avoids the need to create new processes for each task, which considerably reduces the fork overhead. In addition, it is useful to initialize global worker-scope variables, as they only need to be initialized once at worker process creation.

We have implemented the \textit{job queue} pattern for the Lithops multiprocessing \texttt{Pool}. In \texttt{Lithops.multiprocessing.Pool}, workers are long-lived functions that are invoked when the \texttt{Pool} object is created. Operations on the pool (\texttt{map()}, \texttt{apply()}) generate Lithops tasks, but instead of invoking new functions to execute those tasks, they are queued in a Redis list. The worker functions pick up and execute tasks from the queue as they are generated. Once the Pool is closed (\texttt{terminate()}), a message is sent to the workers to terminate their execution.

The main advantage of this implementation is that the overhead of submitting a set of tasks to a Redis list is much lower than invoking a function for every task. With Redis, we can submit all tasks at once with a single \texttt{LPUSH} command, while invoking functions is sequential and the overhead depends on the API and architecture of each FaaS service. Also, reusing functions to execute multiple tasks avoids stragglers caused by cold invocations. However, the main drawback is the function execution time limit. Although the time limit has been increasing over the years (for example, AWS Lambda now supports invocations of up to $15$ minutes \cite{awslambdadocu}), this limit prevents running longer executions.

%
%

\subsection{Disaggregated memory resources}
\label{sec:multiproc}

We have chosen Redis for its simplicity of deployment, in-memory storage and high performance. Redis differs from other traditional key-value databases in that the value has a type, such as \texttt{LIST}, \texttt{STRING}, \texttt{HASHSET}, etc. The different operations available on these data types facilitate the implementation of some of the communication and synchronization abstractions present in Python's \texttt{multiprocessing} library. A single node Redis deployment guarantees consistency and the correct order of read and writes, since Redis is single-threaded, and data is backed in disk for restart recovery, but node failure tolerance is not provided.

Serverless functions are not addressable and lack the ability to open a connection to one from another. To circumvent this limitation, many works have explored using TCP connections through an addressable external proxy residing in a VM \cite{excamera, infinicache}, or using NAT hole-punch techniques to break FaaS networking and enable a P2P-like connection between functions \cite{boxer}. However, all these approaches still need an external rendezvous server, but most importantly, they don't support collective communications. Many multiprocessing abstractions require synchronized and atomic access, such as queuing an item to a \texttt{Queue} or signaling the release of a \texttt{Lock}, and Redis single-threaded implementation meets this requirements in a safe but fast manner.

In Python's \texttt{multiprocessing}, processes need a reference to the objects that represent a shared state resource (such as a \texttt{Queue} or \texttt{Manager} instance). The parent process creates all resources and then it passes a reference to the child processes when they are forked. With Lithops, objects and references passed to functions have to be serializable. To maintain the same behavior, we have followed a pattern in which each resource object (\texttt{Queue}, \texttt{Pipe}...) acts as a proxy to the key-value pair in Redis, which is where the state resides. Each object is uniquely identified and corresponds to a specific Redis key-value pair. Each proxy resource implements reference counting for garbage collection. The counter is consistently stored in Redis, and the resource is deleted from Redis when references reach zero. In addition, each resource incorporates a key expiration time of an hour by default. It is used as a backup to in case there is an error in the code and the program terminates abruptly, since the reference counting mechanism might not delete the resource in a graceful manner.

The different abstractions available and a brief description of how they have been implemented are listed below:

\noindent\textbf{Message passing:} \textit{Pipes} are used for duplex communication between a pair of processes. A new \texttt{Pipe} returns a tuple of two \texttt{Connection} objects, each  corresponding to a Redis \texttt{LIST}. Data can be written to one end of the \texttt{Pipe} using \texttt{Connect\-ion.send()} and it can be read using \texttt{Connection.recv()} on the other end. The \texttt{send()} method executes an \texttt{LPUSH} command to put data in the tail of the list, and the \texttt{recv()} method executes an \texttt{BLPOP} command to get data from the head list. This way, the list is treated as a FIFO queue. The \texttt{BLPOP} gets and removes the head item of the list, or blocks until there is an element available. \textit{Queues} are implemented the same way as \texttt{Pipes}, the difference being that more than two processes can put or remove items from a queue. Redis maintains the order of puts and gets consistent.

\noindent\textbf{Shared state:} \textit{Array} and \textit{Value} are used to share memory in Python multiprocessing. Only basic C type values can be put into the array or value. They are implemented using the \texttt{LIST} type. Processes can read and write to specific indexes or slices of the array. A \texttt{Value} is an \texttt{Array} of size 1. We have opted for using the \texttt{LIST} type instead of \texttt{STRING} because \texttt{STRING} values are limited to 512 MB in size, while in lists, each element of the list can will be at most \texttt{sizeof(long double)} in size, and lists can hold up to $2^{32} - 1$ elements.

\texttt{Managers} allows to create Python resources in a separate process, and they are accessed via sockets. Managers are used to share a basic Python data type with multiple processes, such as a \texttt{dict} or \texttt{list}. The implementation of those types is trivial using Redis, since it provides \texttt{HASHSET} and \texttt{LIST} types natively. A \texttt{Manager} also permits the creation of user-defined classes, which reside instantiated in the Manager and other processes use Remote Method Invocation to access it. To provide a similar behavior using Redis, we have made each process have a local instance of the \texttt{Manager} class, but the state of the user-defined class instance (i.e. its attributes) is remotely stored in Redis as simple key-value pairs. A \texttt{Lock} ensures that attributes are accessed by only one process at a time.

\noindent\textbf{Synchronization:} \textit{Semaphores} and \textit{locks} are implemented using the \texttt{LIST} type. When a \texttt{Semaphore} is created, \textit{N} tokens are added to the list, being \textit{N} the initial value of the semaphore. Every \texttt{acquire()} of the semaphore will execute a \texttt{BLPOP} command, which removes a token from the list. The \texttt{release()} method puts the token back to the list with a \texttt{LPUSH} command. If there are no tokens when a \texttt{acquire()} is called (meaning that N processes are currently in the critical section), the \texttt{BLPOP} command will block until other process performs a \texttt{release()} and puts back a token into the list. Note that in this implementation, the value of the semaphore will always be greater than zero. A \texttt{Lock} is a generalization of a semaphore where \textit{N} is $1$.

To implement \textit{Conditions}, multiple \textit{notification lists} are used to notify blocked process awaiting for the condition event. When a process reaches the condition and hangs on the \texttt{wait()} method, the process registers a new list to the \textit{notification list set} and blocks to it with a \texttt{BLPOP} command. The process that satisfies the condition will add an element to each list of the \textit{notification list set}, so that all waiting processes are unblocked and the execution resumes. \texttt{Barriers} and \texttt{Events} are specific cases of \texttt{Condition}.


\subsection{Disaggregated storage resources}

Lithops multiprocessing also implements a replica of \linebreak Python's built-in \texttt{open} function and the \texttt{os.path} module which allows to transparently read and write files and directories stored on disaggregated storage services (like S3) as if it were a local file system. This is especially useful for \textit{FaaS} since the volume that is mounted in the function container is volatile and the data stored there is lost when the execution finishes. In this way, we offer serverless processes a transparent way to save or recover their state. However, it should be noted that, since we are working on immutable data, it is not possible to modify or expand a file as would be done in a traditional network file system without having to rewrite the entire file, which can be problematic for large files. However, as seen in Section \ref{sec:disk_validation}, disaggregated storage services provide much higher parallel read and write throughput than traditional disks used in monolithic machines. Applications that require reading lots of data in parallel (for example, video encoding) can benefit from disaggregated storage to achieve lower execution times.

\section{Evaluation settings}

This section aims to describe the configuration with which the experiments described below have been carried out.



Unless otherwise stated, the experiments have been run with the following settings: Lithops orchestrator runs on a \linebreak m5.2xlarge EC2 host with Ubuntu 20.04, Lambdas use a containerized Python 3.8 runtime with 1769 MB of RAM \footnote{According to AWS documentation \cite{awslambdadocu}, a runtime of 1769 MB of memory is assigned a whole vCPU, being a vCPU a thread of a CPU with Hyper-Threading \cite{awsec2docu}.} as serverless function and Redis 6.2 instance runs on the host machine with Docker.

The host machine and the AWS Lambdas are in the same VPC private subnet, region and availability zone (us-east-1 A), so traffic does not go through a NAT gateway nor the public internet. In case of using S3 as storage backend (will be stated), the S3 bucket is located in the same region and access to S3 is done via a private endpoint. All Lambda functions have been executed using warm containers.

All local monolithic executions have been carried out using on-demand AWS EC2 instances with different number of vCPUs. In particular, we have used the following instances: c5.4xlarge with 16 vCPUs, c5.9xlarge with 32 vCPUs, \linebreak c5.18xlarge with 64 vCPUs and c5.24xlarge with 96 vCPUs. All of these c5 EC2 instances are running Ubuntu 20.04 and are located in the us-east-1 region. Note that the shared memory abstractions for local executions use local memory, while remote serverless processes use Redis.

Since we want to study access transparency, the code used for both configurations is exactly the same, except that we \linebreak change the import statement so that serverless executions use remote stateful abstractions (from \texttt{import multiprocessing} to \texttt{import lithops.multiprocessing}). In other words, all local executions with a single virtual machine make use of the standard Python multiprocessing library and local resources \linebreak (processes, shared memory), while serverless executions use the Lithops multiprocessing library and remote resources instead (serverless functions, Redis).

We have discarded to perform executions in an on-premise environment for two main reasons. First, we can not emulate a Cloud environment using physical resources, since we do not certainly know the hardware used in EC2 and because of Cloud multi-tenant resource-sharing. Second, AWS Lambda runs on EC2 type instances, so comparing local on-premises executions with Lambda would be unfair.

The source code for the different validations is open source and publicly available on GitHub\footnote{\href{https://github.com/aitorarjona/lithops-transparency-validation}{https://github.com/aitorarjona/lithops-transparency-validation}}.

\section{Micro-benchmark evaluation}
\label{sec:microbenchmarks}

The objective of this section is to perform micro-benchmarks that allow us to identify potential overheads and where are they generated, which will help us understand the validation results of the real applications.

\subsection{Fork-join overhead}
\label{sec:overheads}

One of the main overheads when doing parallel computing is the cost of creating a new thread or process. The purpose of this experiment is to measure the overheads generated when invoking multiple parallel serverless processes and analyze how they scale when the number of functions is increased. Lithops allows the usage of different storage backends and monitoring systems for serverless functions. In this experiment, we will compare the performance between using S3 or Redis as storage backend and task monitoring.

The experiment consists of performing a multiprocessing \texttt{Pool} \textit{map} of several sleep functions, each one would sleep for $5$ seconds.

\begin{figure}[!htb]
	\centering
	\includegraphics[width=0.45\textwidth]{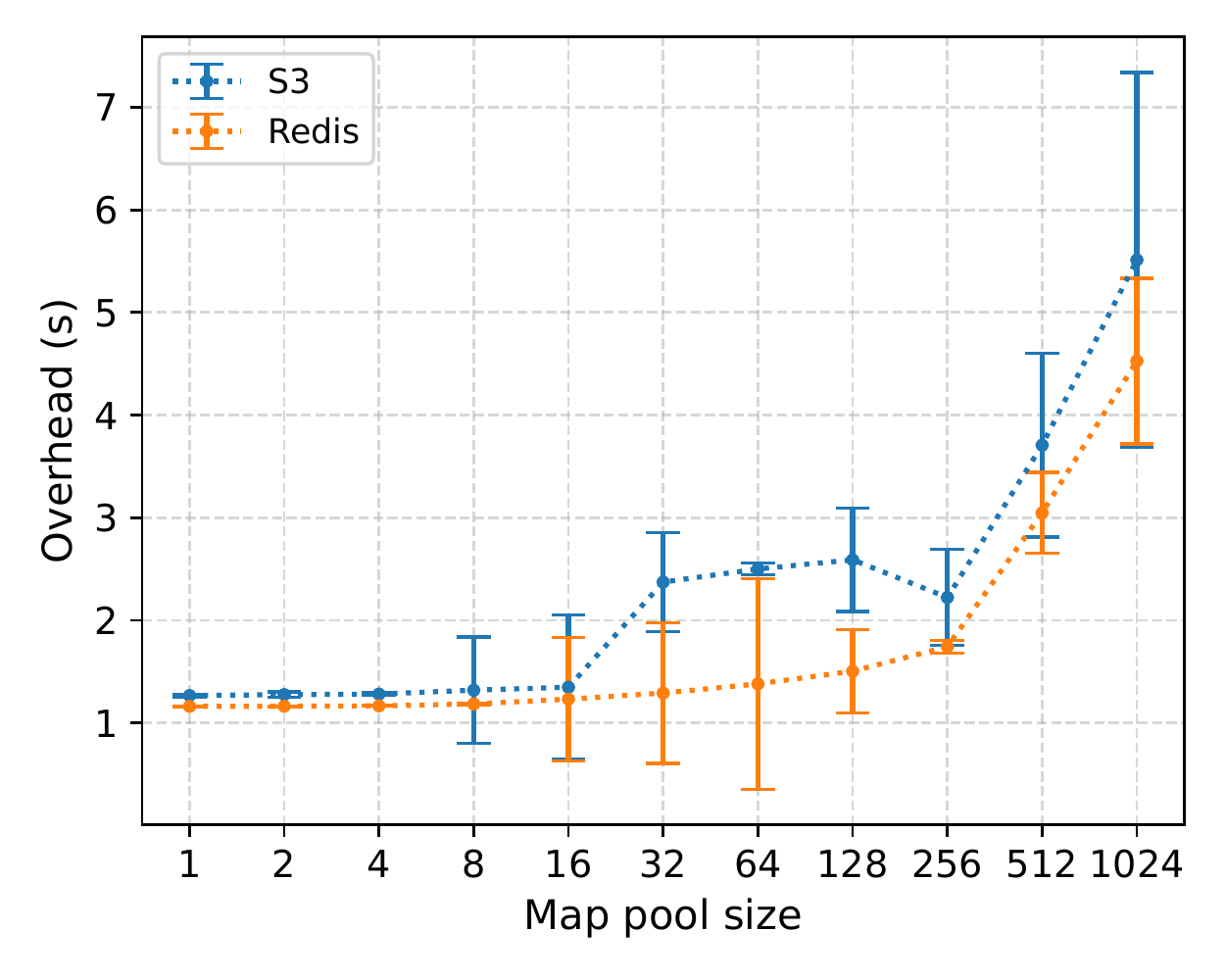}
	\caption{Processes pool invocation results.}
	\label{fig:fork_scale}
\end{figure}

Figure \ref{fig:fork_scale} represents the overall overheads for different parallel executions. The overhead time is calculated by subtracting the sleep time from the total execution time. For low parallelism, both Redis and S3 synchronization provide an overhead of $\approx1$ second. From 32 processes and up, Redis can provide lower overheads than S3: At $1024$ functions, Redis synchronization provides an overhead of $\approx1$ second lower than S3.

We also have measured where these overheads are generated by Lithops and AWS Lambda invocation.

\begin{figure*}[!htb]
	\centering
	\includegraphics[width=0.95\textwidth]{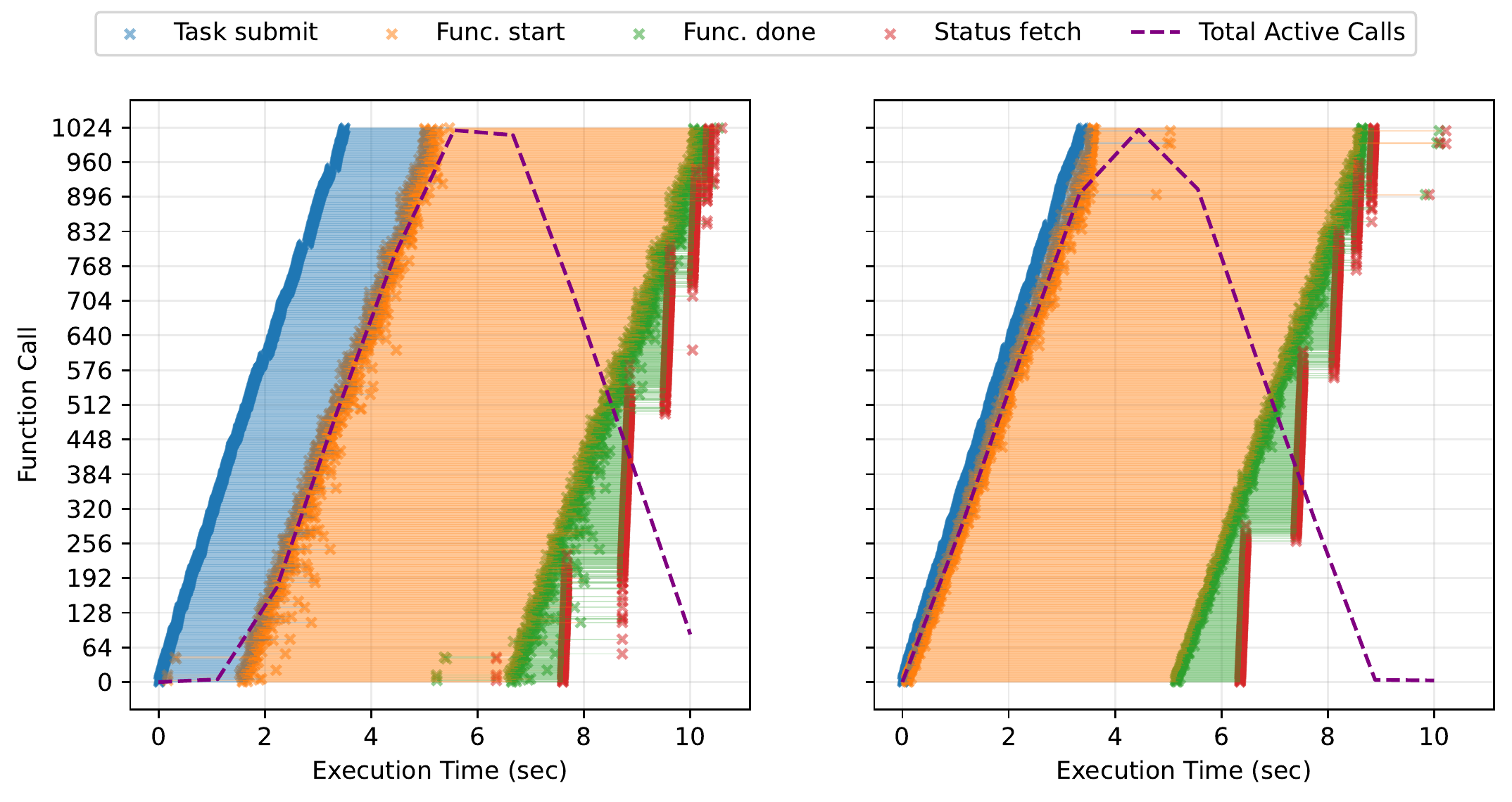}
	\caption{Breakdown of the function's stages for the \texttt{sleep(5)} \textit{map} job. \textbf{(a)} Cold invocation \textbf{(b)} Warm invocation.}
	\label{fig:histogram}
\end{figure*}

\begin{figure*}[!htb]
	\centering
	\includegraphics[width=0.65\textwidth]{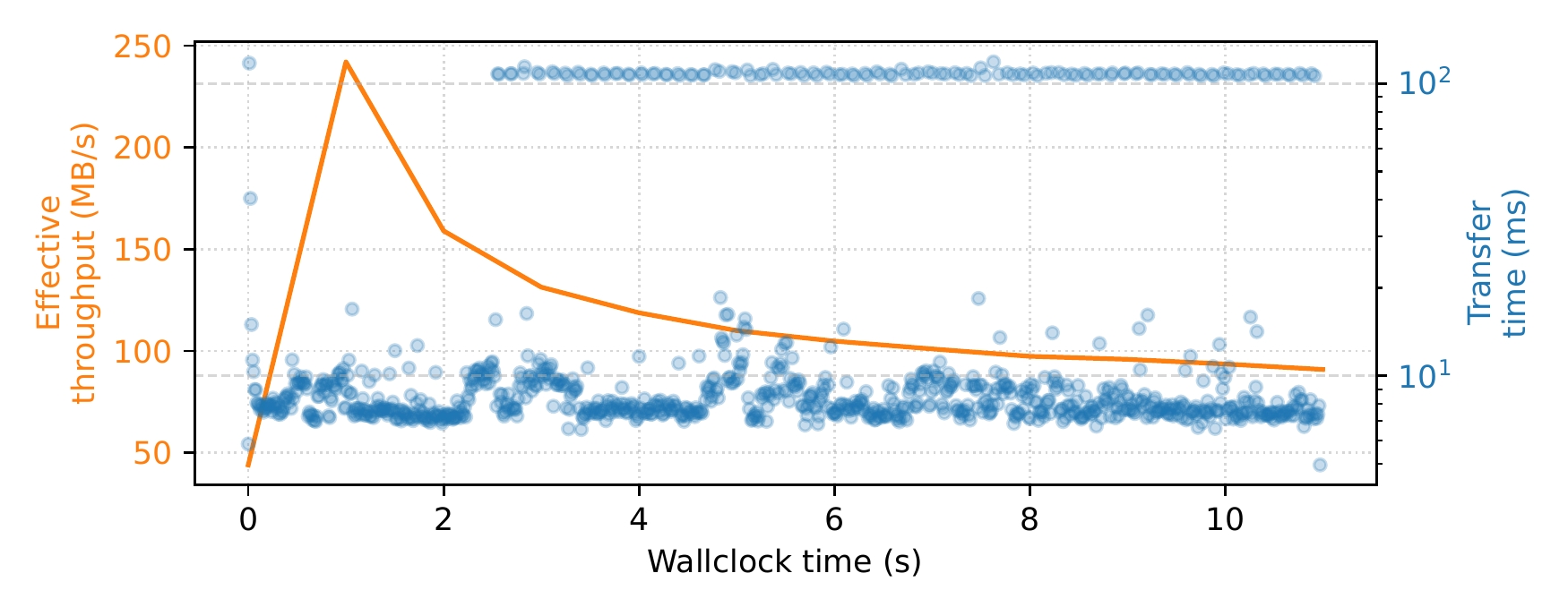}
	\caption{Throughput results.}
	\label{fig:pipe}
\end{figure*}

Figure \ref{fig:histogram} represents a breakdown of a map job of $5$ sleep seconds for $1024$ parallel functions using Redis as storage backend. The execution shown in the left chart used cold containers, while the one on the right used warm containers. We can see the difference in the function start-up overhead. When using cold containers, the overhead is higher because the provider has to allocate resources to run the functions, while when using warm containers, the resources are already allocated and the container is already up and running, so the overhead is much lower. Specifically, warm invocations typically have an overhead of around $200$ ms, while cold invocations can have a more variable and higher overhead. In this case, the overhead of cold invocations exceeds the second and the average overhead obtained is $1.7$ s. We can also see that the start of execution is not instantaneous but linear. This is because asynchronous invocation using Python threads is performed sequentially. This implies that the greater the number of functions, the greater the invocation overhead. It also implies that full parallelism is not achieved immediately. Applications that require exact parallel execution should use some synchronization mechanism (e.g. a barrier).

\begin{table}[!htb]
	\centering
	\begin{tabular}{@{} p{4.5cm} p{1.5cm} p{1.5cm} @{}}
		\toprule
		                                   & \multicolumn{2}{l}{\textbf{\emph{Invocation type}}} \\
		\textbf{\emph{Phase}}              & Cold    & Warm                                      \\ \midrule
		\emph{Serialize data and function} & 0.004 s & 0.004 s                                   \\
		\emph{Upload dependencies}         & 0.002 s & 0.001 s                                   \\
		\emph{Invoke}                      & 1.719 s & 0.258 s                                   \\
		\emph{Function setup}              & 0.052 s & 0.046 s                                   \\
		\emph{Join}                        & 0.628 s & 0.630 s                                   \\ \midrule
		\textbf{\emph{Total}}              & 2.407 s & 0.939 s                                   \\ \bottomrule\hline
		\vspace{0pt}                       &         &
	\end{tabular}
	\caption{Decomposition of generated overheads. The values indicate the average across all functions of a map job.}
	\label{tab:overheads}
\end{table}

Table \ref{tab:overheads} shows the decomposition of the overhead introduced by Lithops and by AWS Lambda. The values indicate the average times of all the functions of the same \textit{map} job. We have differentiated two executions, using warm and cold containers. ``\emph{Serialize data and function}'' and ```\emph{Upload dependencies}'' indicates the time spent to serialize and upload the function data and its dependencies, respectively. These values are constant since both use the same data and storage backend. The ``\emph{Invoke}'' row indicates the time elapsed since the function is invoked until the function begins its execution, and ``\emph{Function setup}'' indicates Lithops worker wrapper setup time. We can see that it is longer for executions using cold containers for the reasons exposed above. The ``\emph{Join}" time indicates the time elapsed since the function ends until it is detected by the Lithops orchestrator. Finally, the sum of all overheads is shown in the ``\emph{Total}'' row.

The overhead time determines the minimum granularity of process tasks. Short-running tasks with run time lower than the overhead will not benefit from distributed serverless execution. Moreover, the closer the granularity is to the overhead time, the more noticeable it will be with respect to the total application execution time.

\subsection{Network latency and throughput}
\label{sec:latency_throughput}

The main limitation and bottleneck for access transparency will be the access latency to remote shared stateful resources. With these experiments, we want to determine the limits of shared memory using Redis and the latency and bandwidth of the network between remote processes.

First, we will determine and compare the latency for local (VM) and remote (serverless functions) communication. We send a variable size payload through a multiprocessing \texttt{Pipe} and measure the round-trip time. Local \texttt{Pipe}s are UNIX pipes while Lithops Multiprocessing \texttt{Pipe} uses Redis Lists. Results on latency are arranged on Table \ref{tab:latency}. We see that the latency of remote communication using Redis is an order of magnitude higher than local communication, so the performance is not comparable. However, we note that for small payloads (less than 1KB), the latency is below a millisecond, which makes the overhead of synchronization operations (like Locks or \linebreak Semaphores) that do not require data passing of minor relevance.

\begin{table}[!htb]
	\centering
	\begin{tabular}{@{} p{4.5cm} p{1.5cm} p{1.5cm} @{}}
		\toprule
		\textbf{\emph{Payload size}} & Remote (Redis) & Local     \\ \midrule
		\emph{1 KB}                  & 0.6 ms         & 0.0463 ms \\
		\emph{1 MB}                  & 23.4 ms        & 2.56 ms   \\
		\emph{100 MB}                & 1.12 s         & 0.288 s   \\ \bottomrule\hline
	\end{tabular}
	\caption{Decomposition of generated overheads. The values indicate the average across all functions of a map job.}
	\label{tab:latency}
\end{table}

Second, we want to measure the maximum throughput of a Lithops multiprocessing \texttt{Pipe} backed by Redis. We send $1000$ messages with a size of 1MB (for a total of 1 GB) through a \texttt{Pipe} to communicate two remote processes running in serverless functions. We can see in Figure \ref{fig:pipe} that the time elapsed for sending a message is stable at $15$ ms, although we can observe some outliers, which could be caused by shared network usage interferences. The total transmission takes $10.5$ seconds, so the effective throughput rate is $\approx90$ MB/s. From this result we can determine the transmission data time between remote processes, which will indicate whether it is worth using disaggregated resources for certain workloads.

%
%
%

\subsection{Computational performance}
\label{sec:computational_performance}
The goal of this experiment is to measure computational performance in an embarrassingly parallel example and to compare the execution time and scalability between a large VM and serverless functions with Lithops.

We used the classic example of the calculation of Pi with the Monte Carlo method to carry out the experiment. In particular, this test is based on sampling \num{3200000000} random points and calculate the number of points that are within the unit circle to extract an approximation of the Pi number. The amount of points to sample is distributed between all the processes, so execution time should decrease when increasing the number of processes.

\begin{figure*}[!htb]
	\centering
	\includegraphics[width=0.95\textwidth]{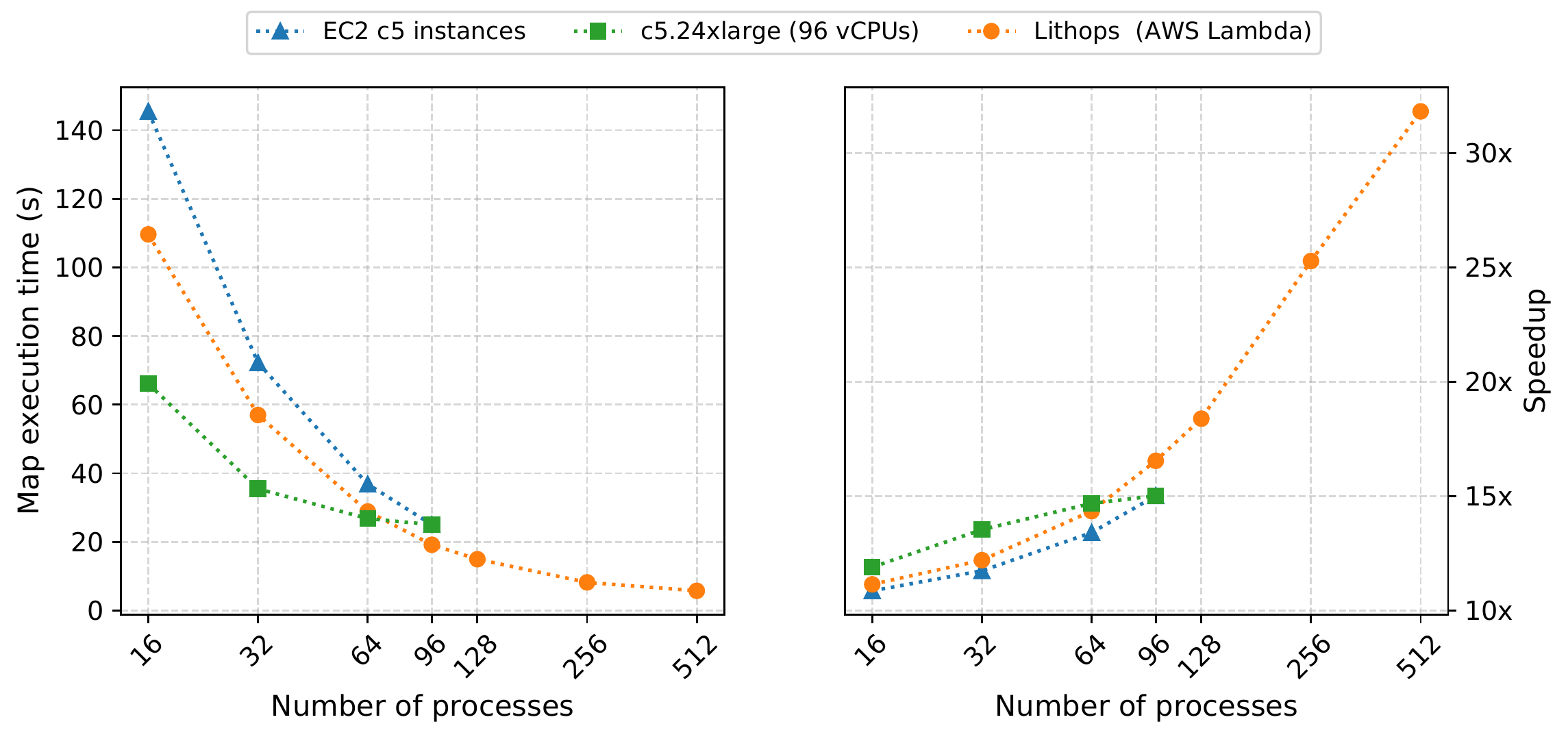}
	\caption{Pi Monte Carlo results.}
	\label{fig:pi_montecarlo}
\end{figure*}

The results of Figure \ref{fig:pi_montecarlo} show that the scalability capacity that can be obtained with Lithops using FaaS goes much further than what a single machine could achieve, despite the executed code is exactly the same. The baseline execution time is $1254.8$ seconds for a single process. For 16 processes, we observe that the performance of the disaggregated system is between $20\%$ and $25\%$ superior compared to the monolithic system. This is caused because, in a VM, the processes share physical CPU cores between them thanks to the use of Hyper-Threading technology, but despite this, the number of floating-point operations is still limited by the physical CPU cores used. On the contrary, functions do not present this limitation, because the physical nodes used by AWS Lambda have Hyper-Threading disabled \cite{firecracker}. However, if we adhere to the documentation, for 1769 MB of memory, the function is allocated one vCPU \cite{awslambdadocu}, i.e. one CPU thread \cite{awsec2docu}. If we refer to the official documentation, we can conclude that, in equivalence of number of vCPUs, functions have a computational advantage over VMs. We can also observe that, for 96 processes, i.e. the VM ceiling, both disaggregated and monolithic systems converge and present approximately the same performance. This is caused because Lithops overheads (see \cref{sec:overheads}) are masked by Hyper-Threading inefficiencies present in VMs.

\subsection{Disk performance}
\label{sec:disk_validation}

The objective of this experiment is to measure the disk read and write capacity and scalability for Lithops' processes running on FaaS, transparently emulating the disk of a VM.

The experiment has two phases: In the first phase, a batch of processes will write a 1GB file to disk. In the second phase, another batch of processes read from disk the data written in the first phase. The experiment is performed for different numbers of processes to study scalability, where we measure aggregate read and write rates. The storage backend used is S3.

\begin{figure}[!htb]
	\centering
	\includegraphics[width=0.45\textwidth]{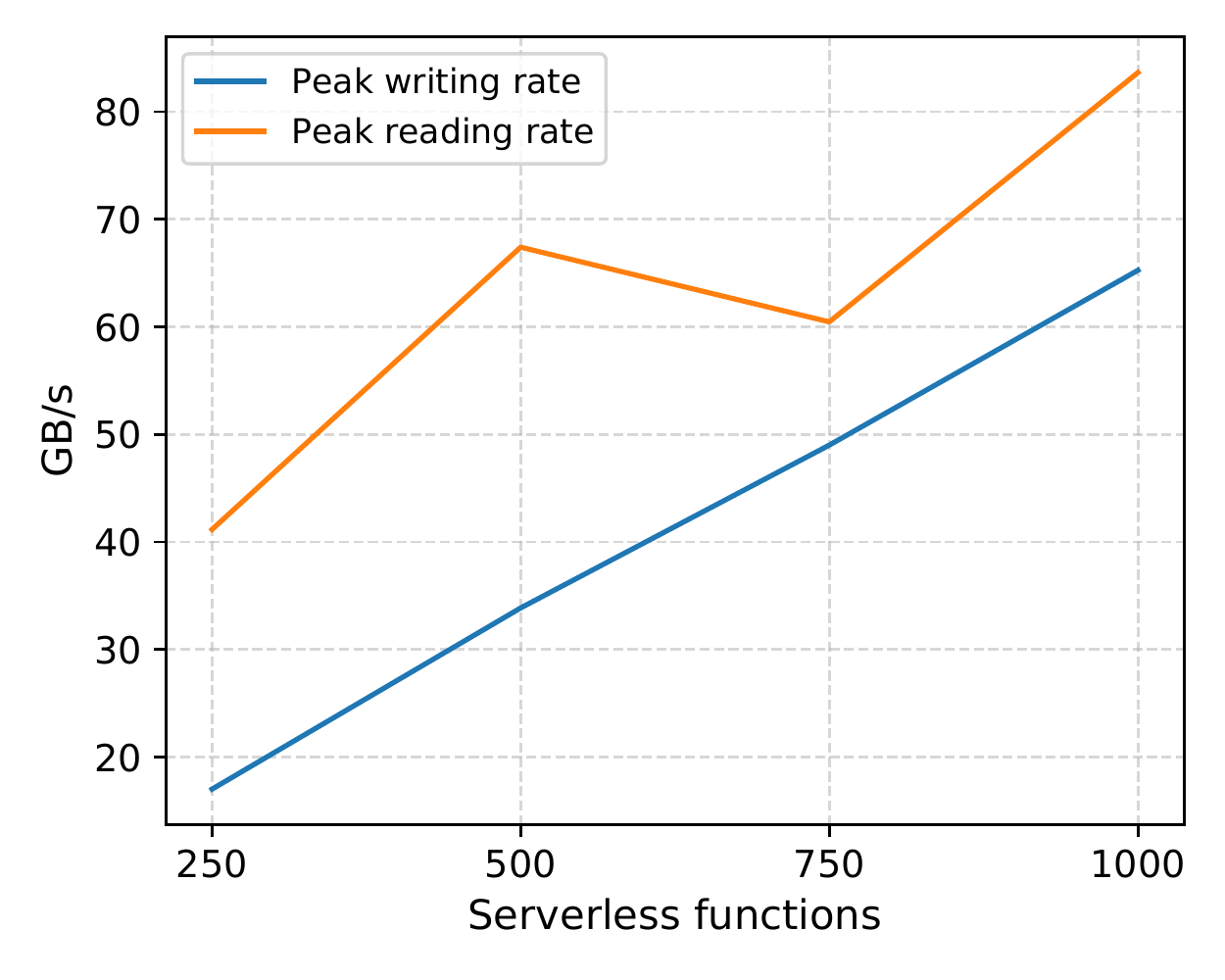}
	\caption{Disk reading and writing rates.}
	\label{fig:disk}
\end{figure}

The results, depicted in Figure \ref{fig:disk}, show high scalability and aggregate bandwidth, with peaks of $80$ GB/s for reads and $65$ GB/s for writes. Compared to General Purpose SSD (gp2) EBS volume, which its maximum throughput is 250 MiB/s\footnote{for volumes bigger than 334 GB regardless of burst credits} \cite{awsebsdocu}, the aggregate write/read throughput of disaggregated storage is considerably higher compared to that of a single volume mount\-ed on a VM. This may be useful for applications that require reading lots of data from storage in parallel since with FaaS we can achieve higher throughput and thus a lower execution time.

\subsection{Shared memory performance}
\label{sec:shared_memory}
We also want to validate the performance loss when using remote shared memory. For this experiment, we have implemented a parallel sorting algorithm using shared memory and compared local and serverless execution. The sorting algorithm consists of splitting in chunks an array, performing quick sort on each chunk in parallel, and then recursively merge pairs of sorted chunks following a tree pattern. This algorithm makes heavy usage of the array as it is iterated over multiple times and the list items are constantly changing positions. We have followed three strategies to implement this algorithm. The first uses multiprocessing's shared \texttt{Array} to store the array and the operations are performed in-place, directly accessing the array indexes. Although the array is stored in shared memory, each process only accesses its corresponding chunk, so there is no need to provide mutual exclusion and critical sections. The second strategy also uses a shared multiprocessing \texttt{Array}, but each process copies its chunk to a local variable, performs the operations on this local slice and then copies back the slice to the shared memory array. The third strategy does not use a shared array, but \texttt{Pipe}s and message passing instead. The parent process sends the chunks to each worker process using pipes, and the workers perform the tree merge phase also passing their chunks using pipes between them.

We have implemented these three strategies to show that, although all three are correct for local execution, the way memory is accessed has a huge impact on the performance when using disaggregated memory.

\begin{table}[!htb]
	\centering
	\begin{tabular}{@{} m{2.4cm} m{1.2cm} m{1.2cm} m{1.2cm} m{1.2cm} @{}}
		\toprule
		\textbf{\emph{Array size}}           &       \multicolumn{2}{c}{5 M}        &       \multicolumn{2}{c}{10 M}       \\ \midrule
		\textbf{\emph{Strategy}}             & \emph{Local} & \emph{Lithops}        & \emph{Local} & \emph{Lithops}        \\ \midrule
		Shared\newline array                 & 23.66 s      & \multicolumn{1}{c}{-} & 79.68 s      & \multicolumn{1}{c}{-} \\
		Shared array\newline with local copy & 15.693 s     & 356.60 s              & 47.11 s      & \multicolumn{1}{c}{-} \\
		Message\newline passing              & 14.27 s      & 17.30 s               & 45.16 s      & 45.63 s               \\ \bottomrule\hline
	\end{tabular}
	\caption{Execution time of the different parallel quick sort implementations with different array size.}
	\label{tab:sharedmem}
\end{table}

Table \ref{tab:sharedmem} shows the execution times of the three alternative implementations using different array sizes ($5$ M and $10$ M) on both local and serverless using 64 processes. For the local execution, we have used a c5 EC2 instance with 64 vCPUs. We can see that Lithops was not able to execute the algorithm using the in-place shared array implementation. This is because each access to a list index is equivalent to a Redis command request. This causes a prohibitive overhead that prevents obtaining competent results using this shared memory approach. The local copy implementation is presented as a low-effort improvement over the in-place shared array implementation. Still, Lithops struggles to perform due to the high data copy overhead. Finally, the message-passing approach using Pipes is presented as the proper way to implement this algorithm using disaggregated memory. Since shared memory is no longer used, Lithops is now capable of providing competent performance compared to local execution. For the $10$ M array size execution, we can see that both local and serverless executions result in the same execution time. We also can see an improvement in the local execution compared to the shared memory alternative. This tells us that, even if we have fast-access local shared memory, it is sometimes not the best alternative even in local executions.

%
%
%

\subsection{Micro-benchmarks conclusion}

As we have seen in the different experiments carried out, the overheads, mainly generated by network latency, are very considerable. We are obtaining overheads of several magnitudes higher (microseconds locally compared to milliseconds remotely). However, the overall performance is not severely affected, because the Hyper-Threading penalty masks the overheads originated by the network. Using FaaS, such as AWS Lambda, gives us great instantaneous scalability, although the additional overheads are more significant. For example, the lack of direct communication between functions makes it necessary to use indirect communication via a disaggregated memory component. In this regard, we are evaluating transparency in a non-optimal scenario -- a better alternative would be to use multiple VMs, where we would benefit from direct communication or where we could avoid the function invocation overhead (with already collocated remote processes). On the other hand, we would lose the ability to scale up quickly and dynamically. Precisely, these attributes are of vital importance when it comes to transparently running Python scripts that use multiprocessing. Until the program is in runtime, we generally cannot know in advance the number of processes or which shared state communication abstractions we are going to use, or whether the number of processes varies throughout program execution. If we were to use a cluster of virtual machines, the total capacity would be fixed, so there would be times when over-provisioning (having more resources deployed than we really need) or under-provisioning (having fewer resources that are throttled by a greater load) would occur.


\section{Applications}
\label{sec:macrobenchmarks}

In this section, we evaluate the behavior of Lithops in four different real use cases in order to test access transparency and to measure performance. The scenarios used are: the implementation of the OpenAI's Proximal Policy Optimization (PPO) algorithm in its Baselines repository, the POET modifications in Evolution Strategies made by the Uber research team, parallel Pandas dataframe transformations using Pandaral·lel, and a hyperparameter tuning using Scikit-learn's Gridsearch running over Joblib. To adapt these applications to serverless, we only had to replace the \texttt{multiprocessing} import with \texttt{Li\-thops.mu\-ltiprocessing}. Since Lithops fully implements the multiprocessing interface, the rest of the code did not need any further modification.

\begin{table}[!htb]
	\centering
	\begin{tabular}{@{} m{1.5cm} m{2cm} m{1.25cm} m{1.25cm} m{1.25cm} @{}}
		\toprule
		\textbf{Application} & \textbf{Algorithm Type} & \textit{Shared State} & \textit{Message Passing} & \textit{Parallel map} \\ \midrule
		Evolution Strategies & Iterative pool map      & \cmark                & \xmark                   & \xmark                \\
		Pandaral·lel         & Scatter-Gather          & \xmark                & \cmark                   & \xmark                \\
		Gridsearch           & Broadcast-Gather        & \xmark                & \xmark                   & \cmark                \\
		PPO                  & Main-\newline Worker    & \xmark                & \cmark                   & \xmark                \\ \bottomrule\hline
	\end{tabular}
	\caption{Summary of algorithm type and stateful abstractions used for every application.}
	\label{tab:apps}
\end{table}

Table \ref{tab:apps} describes the applications and the type of algorithm used in them. We also specify what kind of stateful abstractions are used. In each section, we go into detail about how each algorithm handles the shared state and message passing.

We have taken the PPO and Evolution Strategies applications from the Fiber \cite{fiber} validation since they are real and complex applications that use the multiprocessing library. However, we believe that the comparison with fiber would not be fair, due to (i) container creation times in Kubernetes and AWS Lambda are not comparable and (ii) Fiber uses direct inter-process communication while Lithops uses disaggregated memory. Moreover, they do not indicate enough parameters to replicate their experiments.

Note that the use of these experiments helps us to have complex scenarios in which to check if access transparency can be achieved, in no case it is intended to study the results of the experiments in the fields of artificial intelligence, machine learning or data analysis. Note that as each application is different, so is its scalability, therefore the number of processes used in each application varies.


\subsection{Evolution strategies} 

In this experiment we have used the Paired Open-Ended Trailblazer (POET) \cite{poet} implementation, which is a large \linebreak Python application with about $4000$ LOC (lines of code) using different \textit{multiprocessing} abstractions like \texttt{Pool} or a shared dictionary from a Manager (\texttt{Manager.dict()}). This algorithm is part of the Evolution Strategies category, in which, evolutions of an initial population are carried out iteratively, and those evolutions are executed in parallel. The objective of this test is to analyze and compare the performance and scalability of Lithops in an iterative algorithm that maintains and uses a shared state between processes.

POET uses a shared noise table which is used to generate randomness in the evolution process. This noise table is originally implemented using shared memory. However, it is initialized when the module is loaded, so it is not using Lithops multiprocessing implementation for shared memory. Instead, since this table is read-only, each function can initialize its noise table independently of the shared memory. The algorithm also uses a shared table of parameters that are modified in each iteration. This shared data structure is implemented as a shared multiprocessing \texttt{Manager.dict()} dictionary. Therefore, there is a certain transmission of data that could imply a significant overhead.

The \texttt{multiprocessing} abstractions used in POET are: one \texttt{Context} set to \textit{spawn} mode, one \texttt{Pool} for tasks executions, one \texttt{Manager} with two \texttt{Dict} that contain the shared stated used by all the worker processes from the Pool. 

Each iteration of the algorithm performs a \texttt{Pool.map()} operation. As the task granularity is small (about 3 seconds), to try to mitigate overheads, we used the optimization of the \texttt{Pool} with job queue explained before. To carry out the measurements we have executed $5$ iterations with $512$ batches per chunk and a batch size of $5$. All local executions have been run on a \texttt{c5.24xlarge} EC2 instance. 

\begin{figure*}[!htb]
	\centering
	\includegraphics[width=0.95\textwidth]{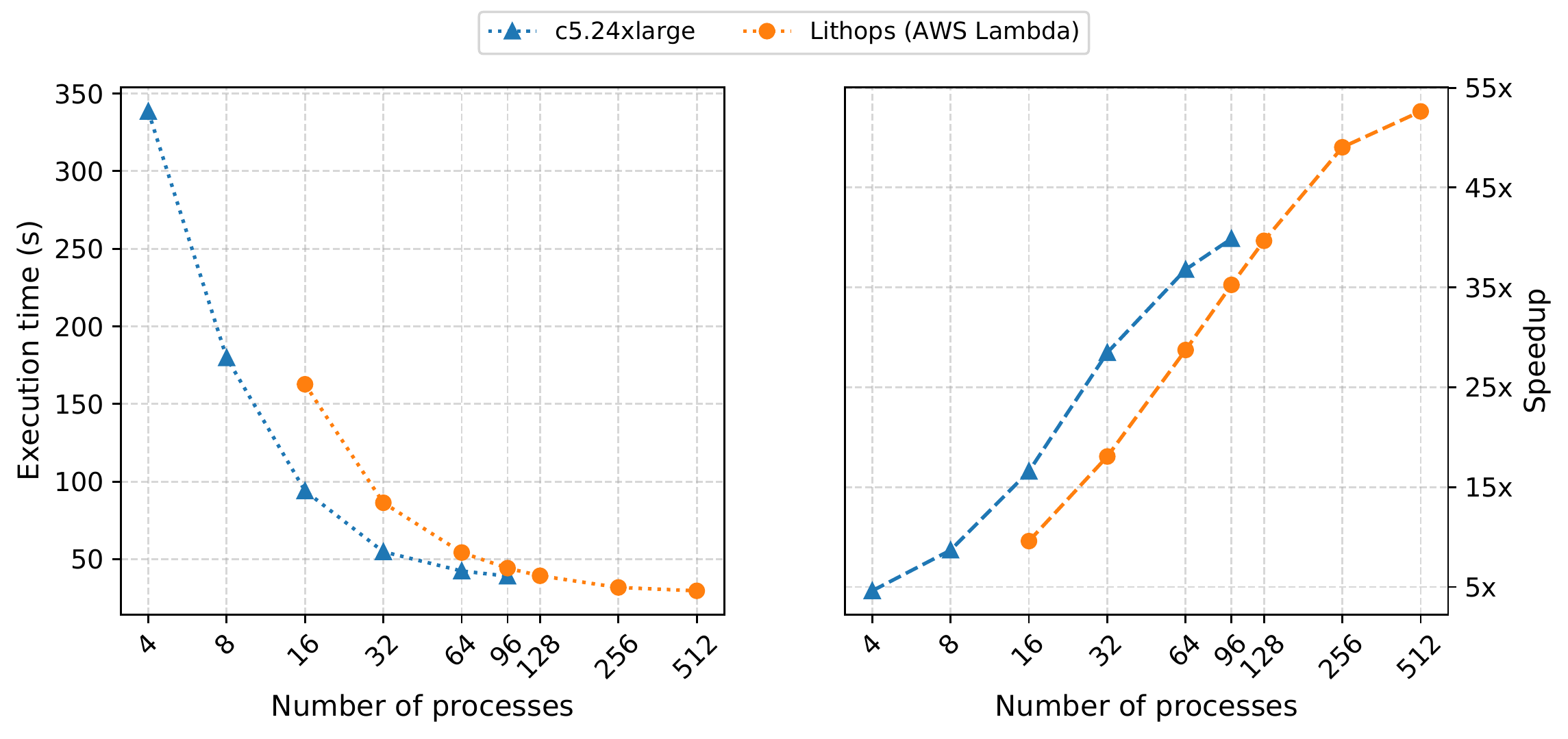}
	\caption{Evolution Strategies execution results.}
	\label{fig:poet}
\end{figure*}

The results in Figure \ref{fig:poet} show that, despite the data transmission and invocation overheads, Lithops maintains constant scalability similar to the scalability of the VM. The maximum speedup of the VM is about $40x$, while Lithops is capable of reaching a speedup of around $53x$, improving the best result of the VM.

\subsection{Pandaral·lel}

Pandas is a Python package, with more than half-million lines of code, that provides data structures to work with relational data. Pandaral·lel \cite{pandarallel} is a Python module with about $1700$ LOC that extends the Pandas functionality adding parallel DataFrames and Series transformations functions like \textit{apply}, \textit{map} and \textit{applymap}. To do so, Pandaral·lel relies on the \texttt{multiprocessing} API, so replacing it by the \texttt{Li\-thops.mul\-ti\-processing} API allows us to execute it in a distributed environment. This experiment aims to measure Lithops' behavior in an embarrassingly parallel task with relatively large data transmissions and analyze how it handles that overhead.

For this experiment we used the Sentiment140 \cite{sentiment140} dataset loaded in a Pandas DataFrame. We used Pandaral·lel \texttt{apply()} function on that DataFrame to perform some sentiment analysis using the textblob Python module. Pandaral·lel first partitions the dataframe according to the number of available workers. Then, it serializes the content of each one and passes it to each function by parameter. Lithops detects the values passed by parameter of each function and transfers them to the storage. The functions are then invoked using \texttt{Pool.map()}. Each function downloads its chunk of dataframe, applies the transformation and, finally, the resulting dataframe is returned to the parent process. During this process Pandaral·lel uses the following \texttt{multiprocessing} artifacts: one \texttt{Context} set to \textit{fork} mode, one \texttt{Manager}, one \texttt{Pool} for tasks execution and one \texttt{Queue} to synchronize the main process with the workers.

\begin{figure*}[!htb]
	\centering
	\includegraphics[width=0.95\textwidth]{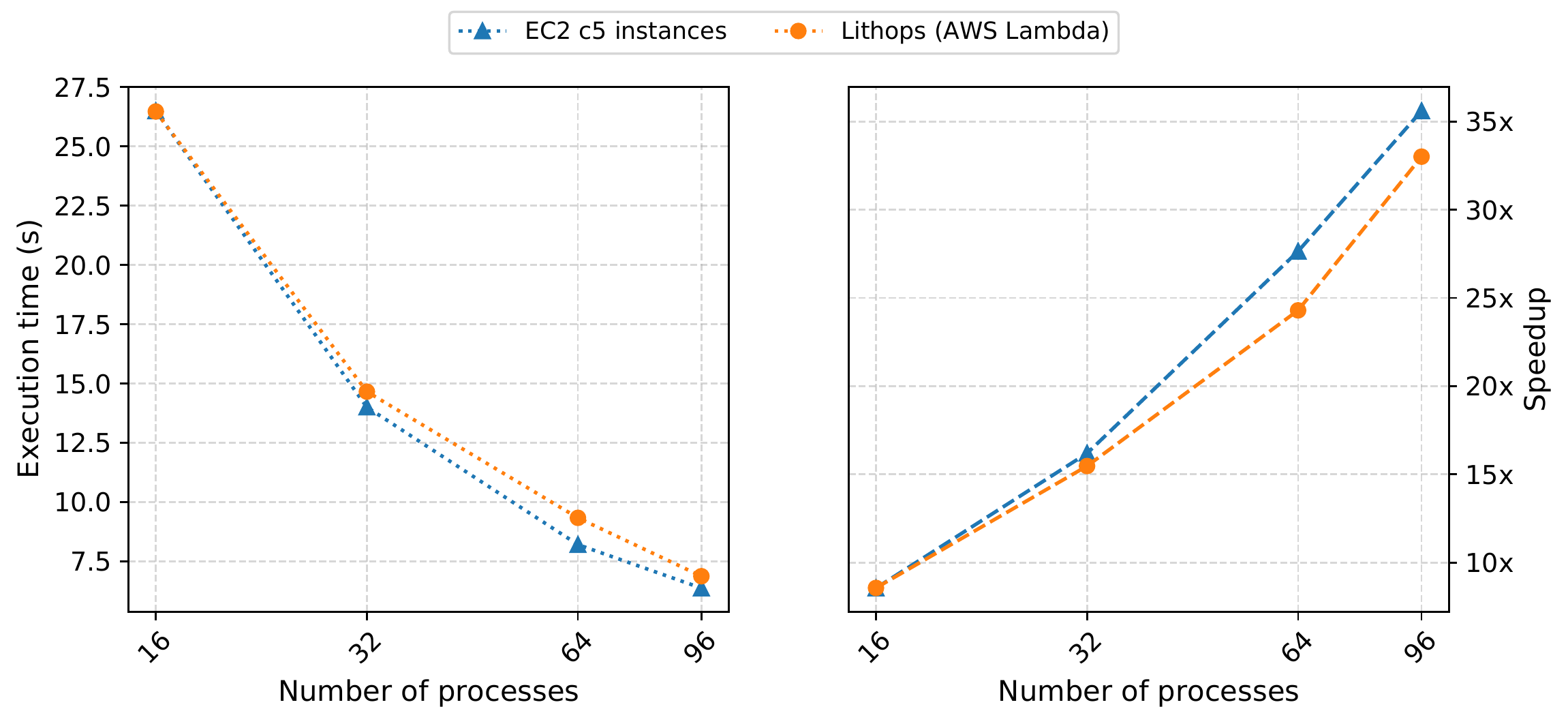}
	\caption{Pandaral·lel execution results.}
	\label{fig:pandarallel}
\end{figure*}

The results in Figure \ref{fig:pandarallel} show that Lithops obtains a $7\%$ lower performance compared to the best result from the VM. It can also be seen that Lithops is capable of maintaining correct scalability up to 96 vCPUs, which can be deployed immediately without previous allocation. 

There are two main reasons for these results. The first reason is data transmission and partition. Notice that the dataset used is above $200$ MB on disc (and about $600$ MB when it is loaded into memory) that has to be transmitted from the Lithops orchestrator entirely to the storage backend.
The second reason is task granularity. In the results can be observed that the differences between VMs and Lithops grow as the number of vCPUs increases. The reason behind this is, as seen in \cref{sec:overheads}, fork-join overhead grows as the number of parallel processes augments. When the granularity of the task is too small (note that with 96 vCPUs the total execution time is below 7 seconds), the Lithops overheads become a significant percentage of the total execution time. For this reason, in 16 or 32 vCPUs the difference between Lithops and VMs is much smaller than when using 64 or 96 vCPUs. 

\subsection{Scikit-learn hyperparameter tuning}

Scikit-learn is a Python module for machine learning built on top of SciPy. It is widely used and considered a standard to build machine learning models to solve classification, regression and clustering problems in the Python ecosystem. Here we want to test Lithops behavior on an embarrassingly parallel scenario with low data transmission.

Some scikit-learn utilities can parallelize their execution via the joblib \cite{joblibdocu} library. Joblib provides a lightweight pipelining in Python, in particular, an API to do easy simple parallel computing. Joblib allows using different parallel backends such as \textit{loky} or \textit{multiprocessing}, but it also allows you to use and develop your own parallel backend. We have created a Lithops backend for joblib using the original \textit{multiprocessing} backend as a template. Changes with respect to the original \textit{multiprocessing} parallel backend are minimal. To use the Lithops backend, the user has to first register this backend into Joblib, and then select it to run the parallel jobs. After that, scikit-learn transparently handles all job lifecycle while the original API calls remain unmodified.

With this new Lithops joblib backend we can execute scikit-learn jobs in serverless functions. In particular, we can use the scikit-learn's Gridsearch over Lithops. The \texttt{model\textunderscore selection} module that implements the \texttt{GridSearchCV} functionality has around $6650$ LOC, which makes it a complex application. In this experiment, we use Gridsearch to do a hyperparameter tuning on a SGD Classifier. We used $30$ MB of an Amazon Reviews dataset to perform a cross-validation of $5$ folds.  Each task requires a chunk of the train and test datasets. We also wanted to compare the behavior of Lithops using two different storage backends (S3 and Redis) so we used them with the default experiment settings explained previously.

\begin{figure*}[!htb]
	\centering
	\includegraphics[width=0.95\textwidth]{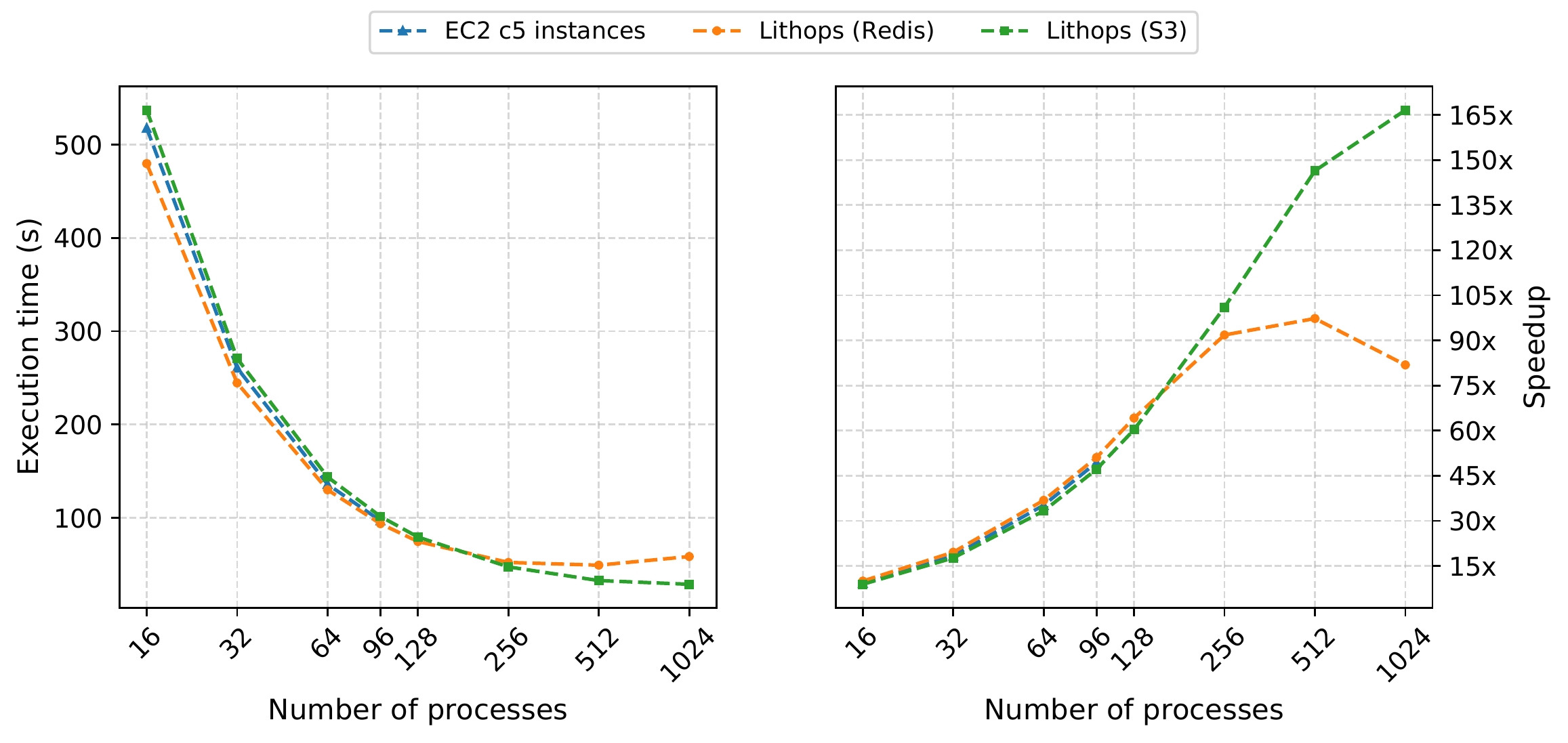}
	\caption{Hyperparameter tuning execution results.}
	\label{fig:gridsearch}
\end{figure*}

The results in Figure \ref{fig:gridsearch} show that with the same number of vCPUs the execution time of the VMs is between $3\%$ and $5\%$ lower than the execution time of Lithops with S3 and between  $3\%$ and $7\%$ higher than the execution time of Lithops using Redis. However, for this type of problem, the scalability of Lithops follows the same progression as the VM and Lithops quickly scale to much higher levels than the VM maximum, obtaining a speedup of up to $3.6x$ times greater. Finally, we can see that Redis is about a $10\%$ better than S3 when few processes are used and, from $256$ processes, Redis begins to saturate but S3 continues scaling correctly. This is caused by the increasing number of concurrent reads of the training and validation data. S3 is able to serve a higher number of concurrent reads compared to Redis which is single-threaded. This implies that S3 has better bandwidth and throughput, which decreases the data read overhead, and consequently decreases the overall execution time.

\subsection{Proximal Policy Optimization} 
\label{sec:ppo}

\begin{figure}[!htb]
	\centering
	\includegraphics[width=0.45\textwidth]{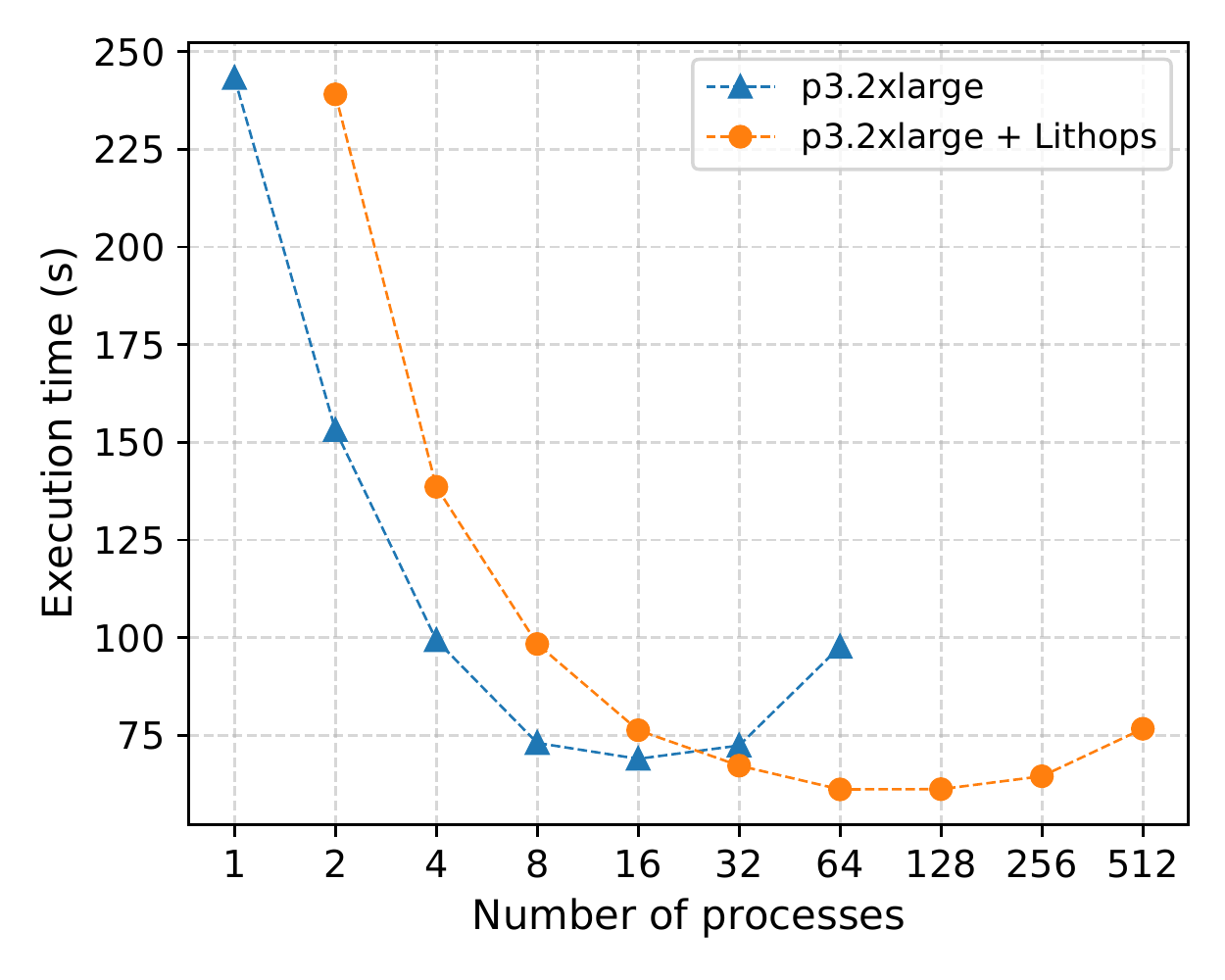}
	\caption{PPO execution results.}
	\label{fig:ppo}
\end{figure}

OpenAI Baselines \cite{baselines} is a set of high-quality implementations of reinforcement learning algorithms. It has been open-sourced to be used as a base, around which, new ideas can be added and as a tool for comparing new approaches in the reinforcement learning field. At the time of writing the paper, the Baselines code repository contains more than $16700$ lines of Python code, so it can be considered as a complex Python module. In this experiment, we want to verify that thanks to the access transparency provided by Lithops \textit{multiprocessing} we can simulate the vertical scaling of a virtual machine using FaaS as processes.

We have used the multiprocessing implementation of the Proximal Policy Optimization (PPO) algorithm from OpenAI baselines. The \textit{multiprocessing} PPO version is the second implementation released by OpenAI, and it inherits some of its structure from the first version, which was based on MPI. For that reason, the \textit{multiprocessing} PPO uses a main\--worker pa\-radigm relying on \texttt{Pipes} for the main to worker communications and vice versa.

The main process is in charge of training the model (a neural network) which, for a given scenario, decides the optimum action to do in order to maximize an objective function. The worker processes are used to simulate the environment in which actions are performed and a reaction is obtained. It is important to notice that each worker process simulates an environment. The training of the model is an iterative procedure where the workers send to the main the actual state of the environment, and it responds with an action to perform in each environment. From the \texttt{multiprocessing} API, PPO uses 1 \texttt{Context} with the \textit{spawn} mode by default. Associated to that \texttt{Context}, it creates 1 \texttt{Process} and 1 \texttt{Pipe} for each environment emulated. The communication between the main and workers (where states and actions are transmitted) is performed using the \texttt{Pipe} associated to each worker \texttt{Process}.

In this experiment, we are training a neural network to play the Atari game Breakout, which is available in the OpenAI GYM \cite{gymOpenAI}. Notice that due to the TensorFlow $1$ dependency, we have used Python $3.7$ in this experiment.


As this algorithm requires the use of a GPU in the main process for the neural network training, the settings for this experiment have been modified. We have used an AWS p3.2xlarge VM as monolithic system, and we tried to scale it vertically using AWS Lambdas. Since the GPU is just used in the main process that runs in the Lithops orchestrator and not in the workers that just do environment simulation, the configuration of the AWS lambdas has not been modified.

The results, available in Figure \ref{fig:ppo}, show that despite the constant communication between processes and the great overhead that this entails, the combination of VM and Lithops achi\-eves a better performance than just the VM. In more detail, the best result for the VM is achieved using 16 processes with a total execution time of $68.92$s and the best result of the VM + Lithops is achieved using 64 processes with a total execution time of $61.10$s, therefore it reduces an $11\%$ the execution time. This validates that we can emulate a vertical scaling of the VM, and that it is possible to add vCPUs to a VM instantly and without prior provisioning thanks to the use of FaaS.



\section{Insights}

After studying the results of the evaluation, we have learned that we are able to transparently move local-parallel applications to distributed settings using serverless.
Nevertheless, we have chosen only four representative Python applications that use the multiprocessing library. By this, we do not want to make the claim that all applications can be transparently scaled using serverless without significant degradation. In this section, we would like to discuss the insights obtained from the evaluation that have implications on the feasibility of transparency using serverless services.

\subsection{Message passing over shared memory}

In distributed systems, a message-passing model is usually used instead of a Distributed Shared Memory (DSM) model. However, there are local-parallel applications that are simply more natural to be developed with shared memory than with message passing. This presents a problem for applications that make heavy usage of local shared memory when transparently moved to a distributed environment, as their performance will be adversely affected, as seen in the experiment of \cref{sec:shared_memory}. If the shared memory access is read-only or with infrequent light writes, it may be possible to implement optimizations that decrease the overhead penalty of accessing DSM.

\subsection{Shared state interfaces}

Clean shared memory abstractions to communicate \linebreak processes are very important. Structured and consistent access to shared state requires suitable programming abstractions. In this case, Python multiprocessing design is a clear facilitator for achieving transparency. The ability to perform parallel execution in Python using threads is limited by the GIL, which prevents multiple threads from running simultaneously on multiprocessor architectures. For this reason, in Python, it is necessary to use processes to have true parallelism. Many of the principles of multiprocessing abstractions, such as Manager, are based on message passing and accessing shared objects (queues, dictionaries, lists\dots) instead of traditional memory sharing. For example, in a multi-thread application written in Java, two parallel threads can access a shared object by a reference pointer. In contrast, in Python multiprocessing, two processes access to shared state by using messages through a third process (the Manager) that has the shared state. The fact that two Python processes can't share the same address space\footnote{except for the multiprocessing.Array abstraction, which can only store basic C types} has facilitated the port of this library to its distributed implementation using disaggregated resources. If the code is not using adequate programming abstractions, full transparency may be impossible.

\subsection{Latencies and overheads}

Overheads are still relevant for many applications. Current Cloud settings still show relevant latency in communications, like hundreds of milliseconds to launch a serverless function, or hundred of microseconds to access in-memory storage services. We have seen that, with equal resources, the overheads generated by creating processes and by the latency of access to shared state are very noticeable. In this line, the granularity of computing tasks is clearly limited by overheads. Very fine-grained computing tasks do not make sense in the current Serverless model, since the overheads can be greater than the task run time.

\begin{table*}[t]
	\centering
	\begin{tabular}{@{} m{3cm} m{1.25cm} m{1.25cm} m{1.25cm} m{1.25cm} m{1.25cm} m{1.25cm} m{1.25cm} m{1.25cm} @{}}
		\toprule
		&               \multicolumn{3}{c}{VM}               &            \multicolumn{3}{c}{Lithops}             & \multicolumn{2}{c}{Comparison} \\ \midrule
		\textbf{\emph{Application}} & \emph{Exec. Time} & \emph{Processes} & \emph{Cost} & \emph{Exec. Time} & \emph{Processes} & \emph{Cost} & \emph{Cost}   & \emph{Time}    \\ \midrule
		Evolution Strategies        & 39.11 s           & 96               & 0.044 \$    & 29.63 s           & 512              & 0.44 \$     & $9.93 \times$ & $1.31 \times$  \\
		Pandaral·lel                & 6.36 s            & 96               & 0.007 \$    & 6.87 s            & 96               & 0.02  \$    & $2.65 \times$ & $0.92 \times$  \\
		Gridsearch                  & 96.97 s           & 96               & 0.109 \$    & 28.73 s           & 1024             & 0.85  \$    & $7.76 \times$ & $3.37 \times$  \\
		PPO                         & 68.92 s           & 16               & 0.058 \$    & 61.10 s           & 64               & 0.11   \$   & $2.82 \times$ & $1.12 \times$  \\ \bottomrule\hline
	\end{tabular}
	\caption{Application's execution time and cost for maximum speedup for AWS ECS and AWS Lambda. ``Comparison'' column indicates how much faster and expensive it is to execute on serverless compared to a virtual machine.}
	\label{tab:cost}
\end{table*}

\subsection{Performance}

Some parallel applications have certain advantages in Cloud Serverless settings that may help to mitigate some of the overheads.

First, Hyper-Threading may cause performance degradation in virtual machines for compute-intensive tasks using all vCPUs. Hyper-Threading makes two threads share some CPU resources like the Arithmetical Logic Unit (ALU). For computationally intensive tasks, two threads are constantly fighting for the shared resources, so the CPU cannot keep up and the execution time is degraded. In High-Performance Computing, disabling Hyper-Threading is a common practice to avoid these problems, although the capacity of effective parallelism is reduced by half. For Serverless Functions, AWS Lambda assigns a vCPU (an Hyper-Threaded CPU thread) per function with a memory configuration of 1769 MB. However, our observations show that the inefficiencies caused by Hyper-Threading in VMs do not occur in Lambda function executions. This provides an opportunity to further improve the parallelism of high-performance applications that require a full physical CPU for better performance.

Second, accessing large volumes of data in Cloud Object Storage from Serverless functions helps to aggregate bandwidth and accelerate data transfers. A single VM cannot compete with parallel data flows from multiple functions.

In addition, as we have seen in the validation of \cref{sec:ppo}, disaggregated resources can serve as ``accelerators'' for a VM. That is, when a VM reaches the maximum occupancy of local resources, it could allocate and move computation to disaggregated resources, e.g., to serverless functions. In this way, we could benefit from both fast-access local memory for shared state-dependent processes running on the VM and high flexibility and scalability for stateless processes.

\subsection{Fault tolerance and serverless services}

The fault tolerance of our solution is based on the assumption that the underlying disaggregated resources are fault tolerant. When programming a monolithic local system, fault tolerance is not taken into account because local resources do not fail. When we move to a distributed environment, if the disaggregated resources (compute, memory and storage) mask the possible failures that may occur, then the application programmer can also assume that they will not fail, and we can continue with the same programming model that does not contemplate error handling and rely on the same local programming model.

Precisely, both AWS Lambda and AWS S3 are fault-tolerant. AWS Lambda can detect and retry failed invocations, while AWS S3 objects are replicated. However, in-memory storage is still not offered as a managed service with scalability and fault tolerance. We are relying on a dedicated Redis service, which must be properly managed to ensure scalability and fault tolerance of the in-memory layer. If the data flows exceed the capacity of this intermediate node, the experiment would fail.

Regarding storage, we are now intercepting file access that is routed to Object Storage. But Object Storage has certain limitations regarding small files or read/write operations. Intensive use of such operations by applications would also preclude transparency. Serverless disaggregated memory and fine-grained storage services are needed in the Cloud.


\subsection{Cost}

Finally, we will discuss the cost of our solution. Nowadays, FaaS is more expensive than IaaS for the same resources (approximately twice as much). The reason is that they are offered as fully managed services that provide fault tolerance and automatic scaling by delegating engineering and management concerns from the user to the Cloud provider, which entails a plus in cost.

In Table \ref{tab:cost} we see a summary of the monetary cost for the execution with the best speedup for both configurations, VM and Lithops, for all applications. The costs have been calculated using the total execution time of the application and the price per hour of the VM in AWS EC2 on demand, and the price per GB/s of AWS Lambda. In the ``Comparison'' column we can see a comparison of both price and time. For example, for the Evolution Strategies application, approximately ten times the cost is required for Lithops to obtain a 30\% improvement in performance compared to a VM. As we can see, our solution clearly does not pay off from a purely monetary perspective.

However, we must state that the cost values only contemplate the cost for the resources used and not the time taken by the application user to manage those resources. The serverless paradigm billing model only charges for the exact resources used by the application execution, contrary to the IaaS billing model where the user is billed by allocated resources, which may be underutilized. For instance, the user may take some time between executions to analyze the results -- time that is wasted by having allocated resources that are not being used at full capacity. We do not want to overemphasize on the cost aspect because it is difficult to calculate the real cost. Data analysts have to cope with an ever-growing availability of tools and services, which prove to be counterproductive because of the management burden. Approximately only a third of developers are actual Cloud experts, so many turn to serverless services because of their ease of adoption, despite the extra cost involved \cite{cloud_future}. In addition, billing models for serverless services may unexpectedly change in the future, so the cost would have to be re-evaluated.

\section{Conclusion}

In this paper, we have demonstrated that Python's multiprocessing message-passing shared state design enables to seamlessly port local-parallel applications over disaggregated serverless resources in the Cloud. Thanks to access transparency, with just changing a single line of code, we are able to deploy a complex local-parallel application in a distributed way in the Cloud. Serverless Cloud services, such as FaaS or Object Storage, allow to massively exploit the parallelism of applications and to further reduce the execution time by increasing parallelism speedup.

We have demonstrated that applications which use stateful abstractions based on message passing, such as queues or pipes, are easy to disaggregate and that the overheads introduced are negligible, obtaining good performance in comparison to the same application running in a big standalone VM. Access transparency is a key to simplify the whole process of moving applications to the Cloud: legacy applications benefit from access transparency since architecture re-engineering would not be required anymore, and data scientists that are familiar with local-parallel programming can instantly and effortlessly scale their code in the Cloud to process bigger workloads.

Nevertheless, performance is severely affected if shared \linebreak memory abstractions are heavily used, since distributed shared memory will never be as fast as local shared memory. In addition, we require programming interfaces where compute resources (such as processes) or state resources (such as queues) are clearly defined.

In conclusion, access transparency is currently possible with some caveats. However, we are optimistic that network latencies will be reduced, and therefore overheads too, so that access transparency will provide ability to program the Cloud \linebreak as a parallel Super-Computer, thus hiding the complexities of distributed systems.

\section*{Acknowledgments}

This work has been partially funded by the EU Horizon $2020$ programme under grant agreement No.$825184$ and by the Spanish Ministry of Science and Innovation and State Research Agency (Agencia Estatal de Investigaci\'on) under grant agreement PID2019-106774RB-C22. Aitor Arjona is a URV Mart\'i Franqu\`es grant fellow.

\bibliography{article}

\begin{thebibliography}{10}
\expandafter\ifx\csname url\endcsname\relax
  \def\url#1{\texttt{#1}}\fi
\expandafter\ifx\csname urlprefix\endcsname\relax\def\urlprefix{URL }\fi
\expandafter\ifx\csname href\endcsname\relax
  \def\href#1#2{#2} \def\path#1{#1}\fi

\bibitem{transparency_coulouris}
G.~Coulouris, J.~Dollimore, T.~Kindberg, G.~Blair, Distributed Systems -
  Concepts and Design (5th edition), Addison-Wesley Longman, Inc., 2012.

\bibitem{rpcs_general_fast}
A.~Kalia, M.~Kaminsky, D.~Andersen,
  \href{https://www.usenix.org/conference/nsdi19/presentation/kalia}{{Datacenter
  RPCs can be General and Fast}}, in: 16th {USENIX} Symposium on Networked
  Systems Design and Implementation ({NSDI} 19), {USENIX} Association, Boston,
  MA, 2019, pp. 1--16.
\newline\urlprefix\url{https://www.usenix.org/conference/nsdi19/presentation/kalia}

\bibitem{time_for_low_latency}
S.~M. Rumble, D.~Ongaro, R.~Stutsman, M.~Rosenblum, J.~K. Ousterhout, It's time
  for low latency, in: Proceedings of the 13th USENIX Conference on Hot Topics
  in Operating Systems, HotOS'13, USENIX Association, USA, 2011, p.~11.

\bibitem{serverless_endgame}
P.~G. L{\'{o}}pez, A.~Slominski, S.~Shillaker, M.~Behrendt, B.~Metzler,
  \href{https://arxiv.org/abs/2006.01251}{Serverless end game: Disaggregation
  enabling transparency}, CoRR abs/2006.01251 (2020).
\newblock \href {http://arxiv.org/abs/2006.01251} {\path{arXiv:2006.01251}}.
\newline\urlprefix\url{https://arxiv.org/abs/2006.01251}

\bibitem{disaggregation}
P.~X. Gao, A.~Narayan, S.~Karandikar, J.~Carreira, S.~Han, R.~Agarwal,
  S.~Ratnasamy, S.~Shenker, Network requirements for resource disaggregation,
  in: 12th USENIX Symposium on Operating Systems Design and Implementation
  (OSDI 16), 2016, pp. 249--264.

\bibitem{disaggregation_and_app}
S.~Angel, M.~Nanavati, S.~Sen,
  \href{http://arxiv.org/abs/1910.13056}{Disaggregation and the application},
  CoRR abs/1910.13056 (2019).
\newblock \href {http://arxiv.org/abs/1910.13056} {\path{arXiv:1910.13056}}.
\newline\urlprefix\url{http://arxiv.org/abs/1910.13056}

\bibitem{pywren}
E.~Jonas, Q.~Pu, S.~Venkataraman, I.~Stoica, B.~Recht, Occupy the cloud:
  Distributed computing for the 99\%, in: Proceedings of the 2017 Symposium on
  Cloud Computing, ACM, 2017, pp. 445--451.

\bibitem{numpywren}
V.~Shankar, K.~Krauth, Q.~Pu, E.~Jonas, S.~Venkataraman, I.~Stoica, B.~Recht,
  J.~Ragan-Kelley, numpywren: serverless linear algebra (2018).
\newblock \href {http://arxiv.org/abs/1810.09679} {\path{arXiv:1810.09679}}.

\bibitem{excamera}
S.~Fouladi, R.~S. Wahby, B.~Shacklett, K.~V. Balasubramaniam, W.~Zeng,
  R.~Bhalerao, A.~Sivaraman, G.~Porter, K.~Winstein,
  \href{https://www.usenix.org/conference/nsdi17/technical-sessions/presentation/fouladi}{{Encoding,
  Fast and Slow: Low-Latency Video Processing Using Thousands of Tiny
  Threads}}, in: 14th {USENIX} Symposium on Networked Systems Design and
  Implementation ({NSDI} 17), {USENIX} Association, Boston, MA, 2017, pp.
  363--376.
\newline\urlprefix\url{https://www.usenix.org/conference/nsdi17/technical-sessions/presentation/fouladi}

\bibitem{dsm_survey}
J.~Protic, M.~Tomasevic, V.~Milutinovic, A survey of distributed shared memory
  systems, in: Proceedings of the Twenty-Eighth Annual Hawaii International
  Conference on System Sciences, Vol.~1, 1995, pp. 74--84 vol.1.
\newblock \href {https://doi.org/10.1109/HICSS.1995.375407}
  {\path{doi:10.1109/HICSS.1995.375407}}.

\bibitem{note_on_distsys}
S.~C. Kendall, J.~Waldo, A.~Wollrath, G.~Wyant, A note on distributed
  computing, Tech. rep., USA (1994).

\bibitem{lithops}
J.~Sampe, M.~Sanchez-Artigas, G.~Vernik, I.~Yehekzel, P.~Garcia-Lopez,
  Outsourcing data processing jobs with lithops, IEEE Transactions on Cloud
  Computing (2021) 1--1\href {https://doi.org/10.1109/TCC.2021.3129000}
  {\path{doi:10.1109/TCC.2021.3129000}}.

\bibitem{legoOS}
Y.~Shan, Y.~Huang, Y.~Chen, Y.~Zhang,
  \href{https://www.usenix.org/conference/osdi18/presentation/shan}{{LegoOS: A
  Disseminated, Distributed {OS} for Hardware Resource Disaggregation}}, in:
  13th {USENIX} Symposium on Operating Systems Design and Implementation
  ({OSDI} 18), {USENIX} Association, Carlsbad, CA, 2018, pp. 69--87.
\newline\urlprefix\url{https://www.usenix.org/conference/osdi18/presentation/shan}

\bibitem{giantVM}
J.~Zhang, Z.~Ding, Y.~Chen, X.~Jia, B.~Yu, Z.~Qi, H.~Guan,
  \href{https://doi.org/10.1145/3381052.3381324}{Giantvm: A type-ii hypervisor
  implementing many-to-one virtualization}, in: Proceedings of the 16th ACM
  SIGPLAN/SIGOPS International Conference on Virtual Execution Environments,
  VEE '20, Association for Computing Machinery, New York, NY, USA, 2020, p.
  30–44.
\newblock \href {https://doi.org/10.1145/3381052.3381324}
  {\path{doi:10.1145/3381052.3381324}}.
\newline\urlprefix\url{https://doi.org/10.1145/3381052.3381324}

\bibitem{spillner2017transformation}
J.~Spillner, Transformation of python applications into function-as-a-service
  deployments (2017).
\newblock \href {http://arxiv.org/abs/1705.08169} {\path{arXiv:1705.08169}}.

\bibitem{fiber}
J.~Zhi, R.~Wang, J.~Clune, K.~O. Stanley, Fiber: A platform for efficient
  development and distributed training for reinforcement learning and
  population-based methods (2020).
\newblock \href {http://arxiv.org/abs/2003.11164} {\path{arXiv:2003.11164}}.

\bibitem{crucial}
D.~Barcelona-Pons, M.~S\'{a}nchez-Artigas, G.~Par\'{\i}s, P.~Sutra,
  P.~Garc\'{\i}a-L\'{o}pez, \href{https://doi.org/10.1145/3361525.3361535}{{On
  the FaaS Track: Building Stateful Distributed Applications with Serverless
  Architectures}}, in: Proceedings of the 20th International Middleware
  Conference, Middleware '19, Association for Computing Machinery, New York,
  NY, USA, 2019, p. 41–54.
\newblock \href {https://doi.org/10.1145/3361525.3361535}
  {\path{doi:10.1145/3361525.3361535}}.
\newline\urlprefix\url{https://doi.org/10.1145/3361525.3361535}

\bibitem{kappa}
W.~Zhang, V.~Fang, A.~Panda, S.~Shenker,
  \href{https://doi.org/10.1145/3419111.3421277}{Kappa: A programming framework
  for serverless computing}, SoCC '20, Association for Computing Machinery, New
  York, NY, USA, 2020, p. 328–343.
\newblock \href {https://doi.org/10.1145/3419111.3421277}
  {\path{doi:10.1145/3419111.3421277}}.
\newline\urlprefix\url{https://doi.org/10.1145/3419111.3421277}

\bibitem{awslambdadocu}
{Amazon Web Services},
  \href{https://docs.aws.amazon.com/lambda/latest/dg/configuration-function-common.html}{{Developer
  Guide for AWS Lambda}} (2021).
\newline\urlprefix\url{https://docs.aws.amazon.com/lambda/latest/dg/configuration-function-common.html}

\bibitem{infinicache}
A.~Wang, J.~Zhang, X.~Ma, A.~Anwar, L.~Rupprecht, D.~Skourtis, V.~Tarasov,
  F.~Yan, Y.~Cheng,
  \href{https://www.usenix.org/conference/fast20/presentation/wang-ao}{{InfiniCache}:
  Exploiting ephemeral serverless functions to build a {Cost-Effective} memory
  cache}, in: 18th USENIX Conference on File and Storage Technologies (FAST
  20), USENIX Association, Santa Clara, CA, 2020, pp. 267--281.
\newline\urlprefix\url{https://www.usenix.org/conference/fast20/presentation/wang-ao}

\bibitem{boxer}
M.~Wawrzoniak, I.~Müller, R.~Fraga Barcelos Paulus~Bruno, G.~Alonso, Boxer:
  Data analytics on network-enabled serverless platforms, 2021-01, 11th Annual
  Conference on Innovative Data Systems Research (CIDR 2021); Conference
  Location: online; Conference Date: January 11-15, 2021; The conference
  lecture was held on January 12, 2021. Due to the Coronavirus (COVID-19) the
  conference was conducted virtually.
\newblock \href {https://doi.org/10.3929/ethz-b-000456492}
  {\path{doi:10.3929/ethz-b-000456492}}.

\bibitem{awsec2docu}
{Amazon Web Services},
  \href{https://docs.aws.amazon.com/AWSEC2/latest/UserGuide/instance-optimize-cpu.html}{{Optimize
  CPU options, Amazon EC2 User Guide for Linux Instances}} (2021).
\newline\urlprefix\url{https://docs.aws.amazon.com/AWSEC2/latest/UserGuide/instance-optimize-cpu.html}

\bibitem{firecracker}
A.~Agache, M.~Brooker, A.~Iordache, A.~Liguori, R.~Neugebauer, P.~Piwonka,
  D.-M. Popa,
  \href{https://www.usenix.org/conference/nsdi20/presentation/agache}{Firecracker:
  Lightweight virtualization for serverless applications}, in: 17th USENIX
  Symposium on Networked Systems Design and Implementation (NSDI 20), USENIX
  Association, Santa Clara, CA, 2020, pp. 419--434.
\newline\urlprefix\url{https://www.usenix.org/conference/nsdi20/presentation/agache}

\bibitem{awsebsdocu}
{Amazon Web Services},
  \href{https://docs.aws.amazon.com/AWSEC2/latest/UserGuide/ebs-volume-types.html}{{Amazon
  EBS volume types}} (2021).
\newline\urlprefix\url{https://docs.aws.amazon.com/AWSEC2/latest/UserGuide/ebs-volume-types.html}

\bibitem{poet}
R.~Wang, J.~Lehman, J.~Clune, K.~Stanley, {Paired Open-Ended Trailblazer
  (POET): Endlessly Generating Increasingly Complex and Diverse Learning
  Environments and Their Solutions}, ArXiv abs/1901.01753 (2019).

\bibitem{pandarallel}
{Manu NALEPA}, \href{https://github.com/nalepae/pandarallel}{Pandarallel}
  (2021).
\newline\urlprefix\url{https://github.com/nalepae/pandarallel}

\bibitem{sentiment140}
A.~Go, R.~Bhayani, L.~Huang,
  \href{http://www.stanford.edu/~alecmgo/papers/TwitterDistantSupervision09.pdf}{{Twitter
  Sentiment Classification using Distant Supervision}}, Processing (2009) 1--6.
\newline\urlprefix\url{http://www.stanford.edu/~alecmgo/papers/TwitterDistantSupervision09.pdf}

\bibitem{joblibdocu}
{Joblib developers}, \href{https://joblib.readthedocs.io/en/latest/}{{Joblib
  documentation}} (2021).
\newline\urlprefix\url{https://joblib.readthedocs.io/en/latest/}

\bibitem{baselines}
P.~Dhariwal, C.~Hesse, O.~Klimov, A.~Nichol, M.~Plappert, A.~Radford,
  J.~Schulman, S.~Sidor, Y.~Wu, P.~Zhokhov, {OpenAI Baselines},
  \url{https://github.com/openai/baselines} (2017).

\bibitem{gymOpenAI}
G.~Brockman, V.~Cheung, L.~Pettersson, J.~Schneider, J.~Schulman, J.~Tang,
  W.~Zaremba, {OpenAI Gym} (2016).
\newblock \href {http://arxiv.org/abs/arXiv:1606.01540}
  {\path{arXiv:arXiv:1606.01540}}.

\bibitem{cloud_future}
D.~Bermbach, A.~Chandra, C.~Krintz, A.~Gokhale, A.~Slominski, L.~Thamsen,
  E.~Cavalcante, T.~Guo, I.~Brandic, R.~Wolski, On the future of cloud
  engineering, in: 2021 IEEE International Conference on Cloud Engineering
  (IC2E), 2021, pp. 264--275.
\newblock \href {https://doi.org/10.1109/IC2E52221.2021.00044}
  {\path{doi:10.1109/IC2E52221.2021.00044}}.

\end{thebibliography}

\end{document}